\documentclass[useAMS,usenatbib,fleqn]{mnras}

\usepackage{graphicx}
\usepackage{graphics}
\usepackage{color}
\usepackage{url}
\usepackage[caption=false]{subfig}
\usepackage{verbatim}
\usepackage{array}
\usepackage{amssymb}
\usepackage{amsmath}
\usepackage{afterpage}
\usepackage{alphalph}

\usepackage{lscape}
\usepackage{pbox}
\usepackage{supertabular}

\usepackage[T1]{fontenc}
\usepackage{ae,aecompl}

\usepackage{orcidlink}

\usepackage{newtxtext,newtxmath}

\newcommand{\swift}{{\it Swift}}

\volume{520}
\pagerange{4356--4369}
\pubyear{2023}

\title[ASAS-SN Bright SN Catalog 2018-2020]{The ASAS-SN Bright Supernova Catalog -- V. 2018-2020}

\author[K.~D.~Neumann et al.]{K.~D.~Neumann \orcidlink{0000-0002-2701-8433}$^{1}$\thanks{neumann.110@osu.edu}, 
T.~W.-S.~Holoien \orcidlink{0000-0001-9206-3460}$^{2}$, 
C.~S.~Kochanek$^{1,3}$, 
K.~Z.~Stanek$^{1,3}$, 
\newauthor
P.~J.~Vallely$^{1}$, 
B.~J.~Shappee \orcidlink{0000-0003-4631-1149}$^{4}$, 
J.~L.~Prieto$^{5,6}$, 
T.~Pessi \orcidlink{0000-0001-6540-0767}$^{5,7}$, 
T.~Jayasinghe \orcidlink{0000-0002-6244-477X}$^{8,1,\dagger}$, 
\newauthor
J.~Brimacombe$^{9}$, 
D.~Bersier \orcidlink{0000-0001-7485-3020}$^{10}$, 
E.~Aydi \orcidlink{0000-0001-8525-3442}$^{11,\dagger}$, 
C.~Basinger$^{1}$, 
J.~F.~Beacom \orcidlink{0000-0002-0005-2631}$^{3,12,1}$, 
\newauthor
S.~Bose \orcidlink{0000-0003-3529-3854}$^{13}$, 
J.~S.~Brown$^{1,14}$, 
P.~Chen \orcidlink{0000-0003-0853-6427}$^{15}$, 
A.~Clocchiatti \orcidlink{0000-0003-3068-4258}$^{16}$, 
D.~D.~Desai \orcidlink{0000-0002-2164-859X}$^{4}$, 
\newauthor
Subo~Dong \orcidlink{0000-0002-1027-0990}$^{13}$, 
E.~Falco \orcidlink{0000-0002-7061-6519}$^{17}$, 
S.~Holmbo \orcidlink{0000-0002-3415-322X}$^{18}$, 
N.~Morrell \orcidlink{0000-0003-2535-3091}$^{19}$, 
J.~V.~Shields \orcidlink{0000-0002-1560-5286}$^{20}$, 
\newauthor
K.~V.~Sokolovsky \orcidlink{0000-0001-5991-6863}$^{11,21}$, 
J.~Strader \orcidlink{0000-0002-1468-9668}$^{11}$, 
M.~D.~Stritzinger \orcidlink{0000-0002-5571-1833}$^{18}$, 
S.~Swihart \orcidlink{0000-0003-1699-8867}$^{22}$, 
\newauthor
T.~A.~Thompson \orcidlink{0000-0003-2377-9574}$^{1,3}$, 
Z.~Way \orcidlink{0000-0003-0179-9662}$^{23}$, 
L.~Aslan$^{2}$, 
D.~W.~Bishop$^{24}$, 
G.~Bock$^{25}$, 
\newauthor
J.~Bradshaw$^{26}$, 
P.~Cacella \orcidlink{0000-0002-3546-3190}$^{27}$, 
N.~Castro-Morales$^{16}$, 
E.~Conseil$^{28}$, 
R.~Cornect$^{29}$, 
\newauthor
I.~Cruz$^{30}$, 
R.~G.~Farfan$^{31,32}$, 
J.~M.~Fernandez$^{33,32}$, 
A.~Gabuya \orcidlink{0000-0003-1541-7557}$^{34}$, 
\newauthor
J.-L.~Gonzalez-Carballo$^{35,32,36}$, 
M.~ R.~Kendurkar$^{37}$, 
S.~Kiyota \orcidlink{0000-0003-2694-7692}$^{38}$, 
R.~A.~Koff$^{39}$,
\newauthor
G.~Krannich$^{40}$, 
P.~Marples$^{41}$, 
G.~Masi$^{42}$, 
L.~A.~G.~Monard$^{43}$, 
J.~A.~Mu\~{n}oz \orcidlink{0000-0001-9833-2959}$^{44,45}$, 
\newauthor
B.~Nicholls$^{46}$, 
R.~S.~Post \orcidlink{0000-0003-3244-0337}$^{47}$, 
Z.~Pujic \orcidlink{0000-0002-2515-7296}$^{48}$, 
G.~Stone \orcidlink{0000-0001-5888-9162}$^{49}$, 
L.~Tomasella \orcidlink{0000-0002-3697-2616}$^{50}$, 
\newauthor
D.~L.~Trappett$^{51}$, 
W.~S.~Wiethoff$^{52}$ \\
\vspace{0.4cm}
\\
\parbox{\textwidth}{
Affiliations are listed at the end of the paper\\
$^{\dagger}$ NASA Hubble Fellow \\
    }
  }

\date{Accepted 2023 January 27. Received 2023 January 27; in original form 2022 October 13}

\begin{document}

  
\maketitle

\begin{abstract}
We catalog the 443 bright supernovae discovered by the All-Sky Automated Survey for Supernovae (ASAS-SN) in $2018-2020$ along with the 519 supernovae recovered by ASAS-SN and 516 additional $m_{peak}\leq18$~mag supernovae missed by ASAS-SN. Our statistical analysis focuses primarily on the 984 supernovae discovered or recovered in ASAS-SN $g$-band observations. The complete sample of 2427 ASAS-SN supernovae includes earlier $V$-band samples and unrecovered supernovae. For each supernova, we identify the host galaxy, its UV to mid-IR photometry, and the supernova's offset from the center of the host. Updated peak magnitudes, redshifts, spectral classifications, and host galaxy identifications supersede earlier results. With the increase of the limiting magnitude to $g\leq18$~mag, the ASAS-SN sample is nearly complete up to $m_{peak}=16.7$~mag and is $90\%$ complete for $m_{peak}\leq17.0$~mag. This is an increase from the $V$-band sample where it was roughly complete up to $m_{peak}=16.2$~mag and $70\%$ complete for $m_{peak}\leq17.0$~mag.
\end{abstract}


\raggedbottom


\section{Introduction}
\label{sec:intro}

Over the past decade, an increasing number of surveys have systematically scanned the sky in search of supernovae and other transient events. The largest contributors for bright transient discoveries are the All-Sky Automated Survey for Supernovae (ASAS-SN\footnote{\url{http://www.astronomy.ohio-state.edu/~assassin/}}; \citealt{shappee14}), the Zwicky Transient Facility (ZTF; \citealt{bellm19}; \citealt{chen20}), and the Asteroid Terrestrial-impact Last Alert System (ATLAS; \citealt{heinze18}; \citealt{tonry18}). Between 2014 and 2022, ASAS-SN was the only survey to observe the entire visible sky. ASAS-SN is limited to bright transients ($g \lesssim 18.5$~mag), giving it lower discovery rates, but this allows high spectroscopic completeness for its discoveries and provides transients that are relatively easy to study across the electromagnetic spectrum.

In addition to studying supernovae (SNe; e.g., \citealt{bose18a,bose19}; \citealt{hoeflich21}; \citealt{chen22}), ASAS-SN obtains data for a broad range of transients, multi-messenger searches, and variable sources. For transients, these include tidal disruption events (TDEs; e.g., \citealt{holoien19c,holoien19b,holoien20}, \citealt{hinkle21}, \citealt{payne22}, recently), novae (e.g., \citealt{kawash21b,kawash22}), dwarf novae (e.g., \citealt{kawash21a}), and changing look or other active galactic nuclei (AGN; e.g., \citealt{neustadt20}, \citealt{hinkle22}, \citealt{holoien22}). There are multi-messenger searches associated with both neutrino (e.g., \citealt{aartsen18}, \citealt{necker22}) and gravitational wave (e.g., \citealt{dejaeger22}) events. ASAS-SN has produced the first all-sky, homogeneously classified sample of variable stars (e.g., \citealt{jayasinghe19,jayasinghe21}) and is working to expand it with the aid of citizen science (\citealt{christy23}). The astronomical community also makes considerable use of the ASAS-SN photometry, particularly through the ASAS-SN Sky Patrol (\citealt{kochanek17}). 

Each of the robotic ASAS-SN units is hosted by Las Cumbres Observatory (\citealt{brown13}) and consists of four 14-cm telescopes, each with a field of view of 4.5 x 4.5 degrees. Starting in 2014, ASAS-SN operated units in both the Northern and Southern hemispheres with one unit, named Brutus, located on Haleakala in Hawaii, and a second unit, named Cassius, located at Cerro Tololo in Chile. Each unit observed in the $V$-band with a limiting magnitude of $V \sim 17$~mag (see \citealt{shappee14}). In 2017, ASAS-SN acquired three more units: Paczynski, also located at Cerro Tololo; Leavitt, located at McDonald Observatory in Texas; and Payne-Gaposchkin, located at Sutherland in South Africa, all of which observe in the $g$-band and have limiting magnitudes of $g \sim 18.5$~mag in optimal conditions. By the end of 2018, ASAS-SN converted the initial two units to observe with $g$-band filters transitioning ASAS-SN completely to $g$-band observations. ASAS-SN is an untargeted survey and in good conditions, ASAS-SN can observe the entire visible sky of approximately 30,000 square degrees in less than one night (\citealt{shappee14}, \citealt{holoien19a}).

All of ASAS-SN's observations are processed automatically and searched in real-time. ASAS-SN publicly announces new discoveries upon first detection where there is no ambiguity, or after follow-up imaging confirms an initially ambiguous source detection. ASAS-SN reports its discoveries to the Transient Name Server (TNS\footnote{\url{https://wis-tns.weizmann.ac.il/}}). Targets are spectroscopically confirmed by the ASAS-SN team and other groups. The untargeted design and high spectroscopic completeness make ASAS-SN ideal for population studies of nearby SNe and their host galaxies (e.g., \citealt{brown19}; Desai et al. \textit{in prep.}).

This paper is the fifth in a series of ASAS-SN supernova catalogs, and it spans the years 2018 to 2020. The previous catalogs are presented in \citet{holoien17a,holoien17b,holoien17c,holoien19a}. We provide information on all SNe discovered and recovered by ASAS-SN along with information on their host galaxies. By recovered SNe, we mean SNe discovered by a group other than ASAS-SN that were later seen independently by the ASAS-SN transient pipeline. Supernovae not discovered or recovered by ASAS-SN are presented alongside recoveries with similar data. We provide information for all bright SNe ($m_{peak} \leq 18$~mag) gathered first from ASAS-SN data then external sources. The data and analysis presented in this catalog are meant to supersede data from ASAS-SN webpages, TNS, and The Astronomers Telegram (ATels\footnote{\url{https://astronomerstelegram.org/}}) relating to discovery and classification of SNe. 

In Section~\ref{sec:sample}, we describe sources of the data for both ASAS-SN SNe and the externally discovered SNe along with any updated measurements and their host identifications. In Section~\ref{sec:analysis}, we discuss the statistics of the SNe and their host galaxies. In Section~\ref{sec:disc}, we summarize and discuss the uses of the catalog. Where needed, we use a standard flat $\Lambda$CDM cosmology with $H_0=69.3$~km~s$^{-1}$~Mpc$^{-1}$, $\Omega_M=0.29$, and $\Omega_{\Lambda}=0.71$.


\section{Data Samples}
\label{sec:sample}

This section outlines the sources of data for the supernovae and their host galaxies. Tables \ref{table:asassn_sne} and \ref{table:other_sne} present data on SNe discovered by ASAS-SN and other groups, and Tables \ref{table:asassn_hosts} and \ref{table:other_hosts} present data on the host galaxies of the SNe discovered by ASAS-SN and other groups, respectively.


\subsection{The ASAS-SN Supernova Sample}
\label{sec:asassn_sample}

Table~\ref{table:asassn_sne} contains information for the 443 supernovae discovered by ASAS-SN over the three years spanning 2018 January 1 to 2020 December 31. All discovery information regarding supernova names, discovery dates, and host galaxy names were compiled through the ASAS-SN website and TNS. The TNS discovery reports are cited in Table~\ref{table:asassn_sne}. In addition to their ASAS-SN names, the table includes their International Astronomical Union (IAU) names.

All ASAS-SN SNe with classification spectra have measured redshifts. When the supernova has a host galaxy with a measured redshift agreeing with the supernova's redshift, the redshift of the host is listed instead. We acquire these spectroscopic host redshifts from the NASA/IPAC Extragalactic Database (NED\footnote{\url{https://ned.ipac.caltech.edu/}}). The redshift used for unclassified ASAS-SN SNe are their host redshift if the identification of the host is unambiguous due to proximity or visual overlap. Hosts galaxies are identified as the nearest galaxy to the supernova. Core-collapse SNe are associated with star formation and Type~Ia SNe are well correlated with the infrared light (e.g., \citealt{cronin21}), so host identification is generally unambiguous. The one clear exception is a Type Ia supernova associated with the cluster Abell 0194 (see below). Phrased another way, for the densities and clustering scales of galaxies (e.g., \citealt{zehavi11}), typical galaxy separations are $\sim 0.8$~Mpc while $\sim 90\%$ of our host identifications are within $10$~kpc. 

Supernova classifications were primarily from TNS classification reports or ATels in the instances where no TNS classification reports could be found. These sources are cited in Table~\ref{table:asassn_sne}. Supernova classifications were generally based on either the Supernova Identification code \citep[SNID;][]{blondin07} or the Generic Classification Tool (GELATO\footnote{\url{gelato.tng.iac.es}}; \citealt{harutyunyan08}). These packages compare the observed spectrum to template spectra to estimate the redshift, type, and approximate age of the supernova. SNe discovered by ASAS-SN that could not be classified or lacked classification spectra are labeled as Type ``unk''. 

Updated redshifts and classifications are included where reexaminations of archival classification spectra disagree with previous reports. These classification spectra are obtained from TNS and the Weizmann Interactive Supernova data REPository \citep[WISEREP;][]{yaron12}. ASASSN-20qc (AT~2020adgm) has been updated from being typed as a CV to a Type~IIn and ASASSN-18yy (SN~2018hts) has an updated redshift. Reviewing the spectra of ASASSN-18cl (AT~2018ts), we agree with \citet{palmerio18} that it could be an AGN, a TDE, or a Type~IIn, so we treat it as having an unknown type (``unk'').

While ASAS-SN astrometry is usually better than $2\farcs0$ for bright sources, it does not perform as well for faint sources given the $7\farcs0$ pixel scale. More precise astrometry is acquired using follow-up images of the ASAS-SN supernova and the astrometry.net package (\citealt{barron08};~\citealt{lang10}). We use {\sc Iraf} \citep[][]{tody86} to measure centroid positions for each supernova. This technique results in positional errors of $<$1\farcs{0}. ASAS-SN collects these follow-up images at the Las Cumbres Observatory using their 1-m telescopes or from amateur collaborators working with ASAS-SN. While we give preliminary coordinates to TNS and their discovery reports, we announce coordinates measured by follow-up images in discovery ATels, and these are the values reported in Table~\ref{table:asassn_sne}. We calculated the offset between a supernova and its host galaxy using host positions, primarily from NED. 

The reported host of the Type~Ia supernova ASASSN-18nt (AT~2018ctv) is the cluster Abell 0194. The closest galaxies to it are Minkowski's Object and NGC 0541 with offsets of 72\arcsec and 124\arcsec, but it is likely an intracluster supernova rather than being associated with either galaxy.

For each supernova, we produced new image subtracted light curves in magnitudes using a reference image excluding any epoch with significant emission from the supernova. We fit the region near its peak with a parabola to determine the peak magnitude. \citet{holoien17a,holoien17b,holoien17c,holoien19a} reported the brighter of the brightest observed magnitude and the parabolic fit. Here we use the peak magnitude from the parabolic fit unless there are too few data to carry out the fit, in which case we simply report the peak observed magnitude. Desai et al. (\textit{in prep.}), in determining the luminosity function for Type Ia SNe in ASAS-SN, found that the procedure used in \citet{holoien17a,holoien17b,holoien17c,holoien19a} tended to overestimate the peak brightness by a median of 0.3 magnitudes simply because the large photometric uncertainties of the faint SNe (i.e. most of them) makes the brightest observed magnitude a biased estimator. Desai et al. (\textit{in prep.}) will include updated $V$-band peak magnitude measurements for all Type~Ia SNe up to the end of 2017. The peak magnitudes for SNe discovered by ASAS-SN from 2018 to 2020 are reported in Table~\ref{table:asassn_sne} separately for the $V$-band and $g$-band although there were $V$-band observations only in 2018.


\subsection{The Non-ASAS-SN Supernova Sample}
\label{sec:other_sample}

Table~\ref{table:other_sne} details all SNe discovered by groups other than ASAS-SN from 2018 to 2020. These external groups include both professional and amateur supernova searches. We include SNe only if they have spectroscopic classifications and peak magnitudes $m_{peak}\leq18$~mag. We based the list on the ``Latest Supernovae'' website\footnote{\url{http://www.rochesterastronomy.org/snimages/}} created by D.~W.~Bishop \citep{galyam13}. This site assembles sources, including ATels and TNS, to build an annual database of SNe. TNS is used to verify data from the Latest Supernovae site rather than as the primary source because some discovery sources do not utilize TNS.

For the Non-ASAS-SN sample, we acquired supernova names, IAU names, discovery dates, coordinates, host galaxy names, peak reported magnitudes, spectral types, redshifts, and discovery sources for each SNe from the Latest Supernova website. We used NED to gather host galaxy coordinates to calculate host offsets from angular separation and host redshifts for more accurate measurements. For the majority of SNe without a reported host galaxy, we used NED to locate the nearest possible host galaxy. To verify the accuracy of these hosts, we compared SNe redshifts to host redshifts as well as angular and physical offsets. When the supernova still lacked a possible host in NED, we used the Pan-STARRS DR2 (\citealt{chambers16}) catalog to identity possible hosts and their coordinates by proximity. All of these hosts were within 2\arcsec or visibly lay on top of the SNe. 


\begin{figure*}
\begin{minipage}{\textwidth}
\centering
\subfloat{{\includegraphics[width=0.95\linewidth]{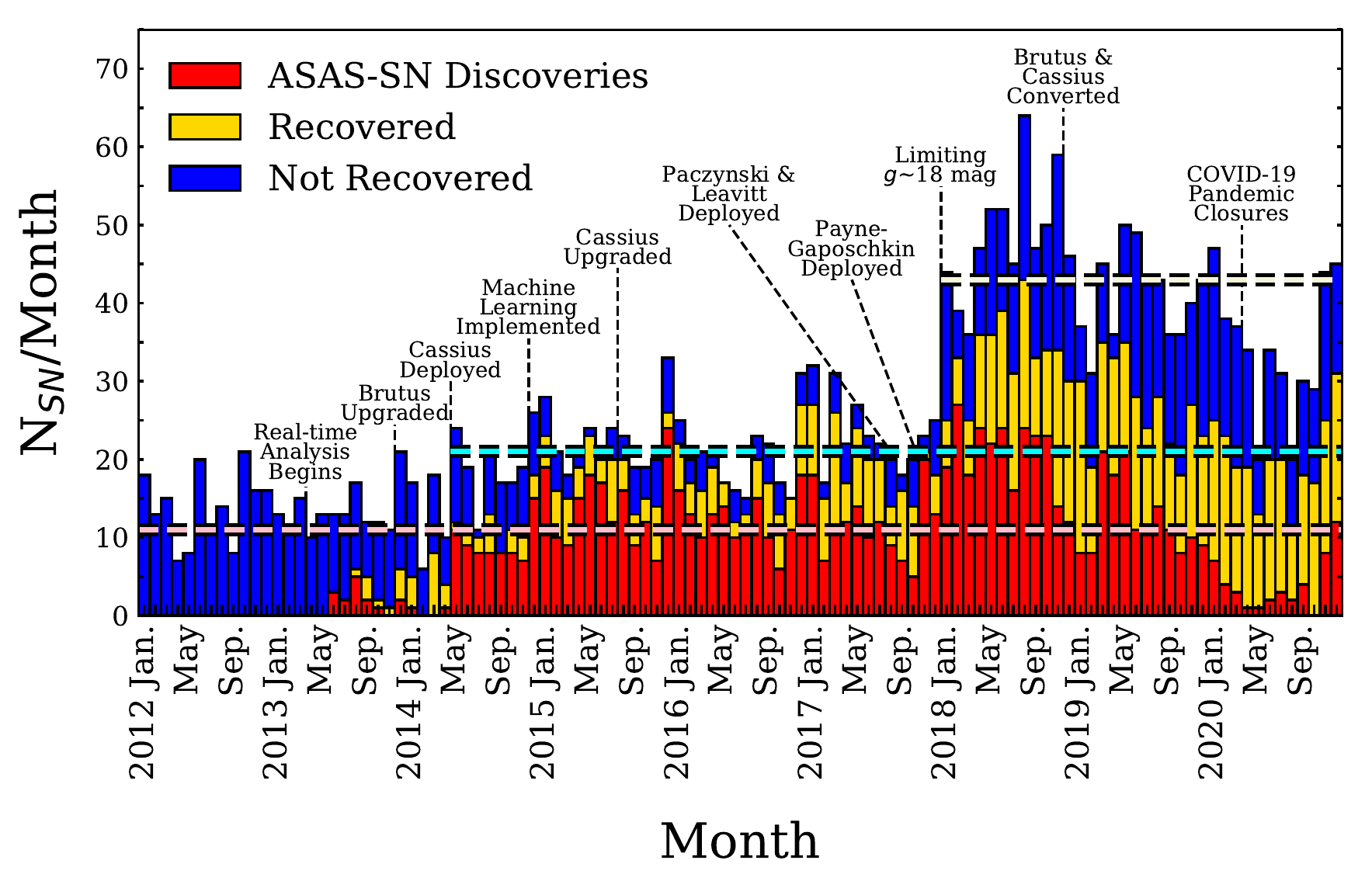}}}

\caption{Histogram of the monthly supernova discoveries from 2012 through 2020. SNe discovered by ASAS-SN are red, SNe recovered but not discovered by ASAS-SN are yellow, and SNe not discovered or recovered by ASAS-SN (or were found prior to ASAS-SN) are blue. Important milestones are labeled. Median discovery rates for 2010 to 2012 ($V\leq17$~mag, pink, before ASAS-SN), 2014 May to 2017 ($V\leq17$~mag, cyan), and 2018 to 2020 ($g\leq18$~mag, beige) are shown by the dashed lines.}

\label{fig:histogram}
\end{minipage}
\end{figure*}

The Latest Supernova website reports maximum magnitudes detected in various filters where this maximum magnitude may not be that of the actual peak of the supernova. To better compare the ASAS-SN and Non-ASAS-SN samples, we again produce ASAS-SN host-subtracted light curves. We used parabolic fits to estimate the peak magnitude and report either this value or the peak measured value if the parabolic fits cannot be done in Table~\ref{table:other_sne}. This was only done for the SNe detected by ASAS-SN. This includes all SNe recovered by ASAS-SN and some of the non-recovered SNe. For ZTF20abqvsik (SN~2020rcq), one of the brightest SNe discovered in 2020, our light curves do not have any data until 3 months after discovery with a $g$-band magnitude of 15.2~mag. The maximum magnitude detected for this supernova is given as 11.8~mag in the $C$-band by Giancarlo Cortini (``Latest Supernovae''). The ZTF $g$-band light curves for this period are not public.

For several Non-ASAS-SN SNe that lacked or had questionable redshifts, we used publicly available spectra from TNS and WISEREP to check the classification and redshift. We reclassified SNe ZTF19aczlqcd and MASTER OT J000256.70+323252.3 (also known as SN~2019wzz and SN~2019el) as an M-dwarf flare and a CV respectively. We have updated redshifts for ATLAS20bfpj (SN~2020aagy), ATLAS20rzv (SN~2020nxt), Gaia20ffa (SN~2020zlz), MASTER OT J005402.48+471051.7 (SN~2018cgq), and ZTF18aavwurv (SN~2020pnn). Additionally, we concur with David Bishop of ``Latest Supernovae'' that AT~2021ekf and the classified supernova 10LYSEnhv are the same object. They were discovered nearly simultaneously, and they have a reported coordinate offset of only $0\farcs05$. 

We report the discovery group for each supernova in Table~\ref{table:other_sne}. Discoveries by non-professional surveys are labeled as ``Amateurs''. The names of the amateurs are included in the complete machine-readable version of Table~\ref{table:other_sne}. Following the pattern seen in the previous ASAS-SN catalog \citep{holoien19a}, amateur discoveries have diminished as the professional surveys have increased in scale. This decrease and the extension to a fainter limiting magnitude dropped amateurs to 5th in number of supernova discoveries in 2018-2020 compared to 3rd in 2017.

Table~\ref{table:other_sne} notes if a Non-ASAS-SN supernova was recovered by ASAS-SN during standard operations. These recovered SNe can be used in any statistical analysis of the ASAS-SN SNe. The missed SNe provide information on the completeness of ASAS-SN.


\begin{figure*}
\begin{minipage}{\textwidth}
\centering

\subfloat{{\includegraphics[width=0.8\textwidth]{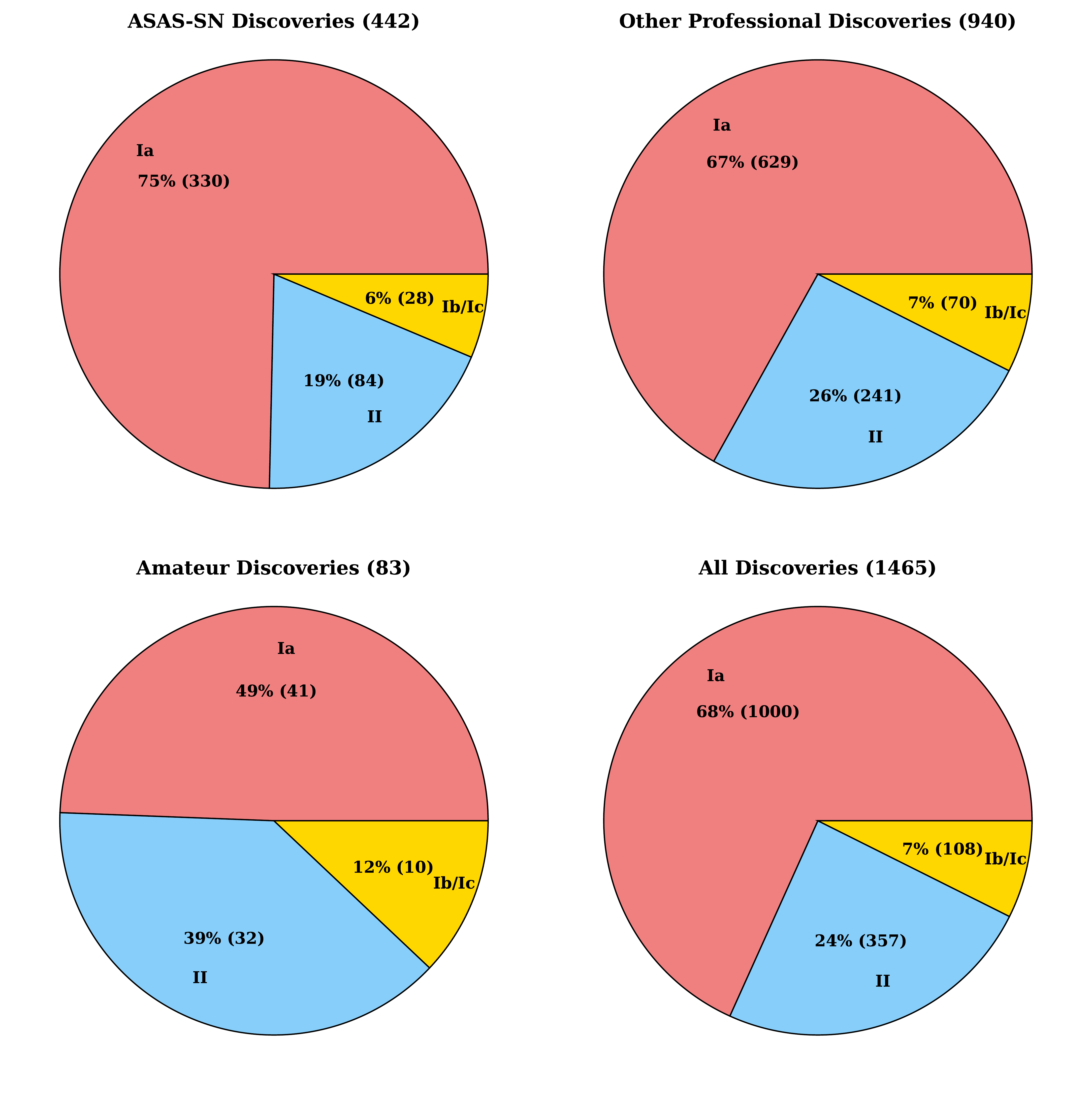}}}
\caption{Classification breakdown for supernova discoveries between 2018 January 1 and 2020 December 31 for those found by ASAS-SN (\emph{Upper Left}), other professional surveys (\emph{Upper Right}), amateurs (\emph{Lower Left}), and all recent SNe (\emph{Lower Right}). We exclude 9 superluminous SNe, 4 incompletely typed SNe, and 142 untyped SNe. Sub-classes are included with their ``parent class'' (e.g., Type~IIn SNe are counted as Type~II SNe).}

\label{fig:piechart}
\end{minipage}
\end{figure*}


\subsection{The Host Galaxy Samples}
\label{sec:host_sample}

In Tables~\ref{table:asassn_hosts} and~\ref{table:other_hosts}, we provide the Galactic extinction estimates towards the host galaxy and host magnitudes from the near-ultraviolet (NUV) through infrared (IR). The Galactic extinctions ($A_V$) are from \citet{schlafly11} using the host coordinates from NED. We collected NUV magnitudes from the Galaxy Evolution Explorer \citep[Galex;][]{morrissey07} All-Sky Imaging Survey (AIS), $u$ magnitudes from the Sloan Digital Sky Survey (SDSS) Data Release 14 \citep[DR14;][]{albareti17}, $grizy$ magnitudes from the Panoramic Survey Telescope \& Rapid Response System \cite[Pan-STARRS;][]{chambers16}, NIR $JHK_S$ magnitudes from the Two-Micron All Sky Survey \citep[2MASS;][]{skrutskie06}, and mid-IR $W1$ and $W2$ magnitudes from the Wide-field Infrared Survey Explorer \citep[AllWISE;][]{wright10}.

For hosts not detected by 2MASS, we set $J$ and $H$ band upper limits based on the faintest galaxies in the $2014-2020$ sample ($J>17.0$~mag, $H>16.4$~mag). For host galaxies detected in WISE $W1$ but not in 2MASS $K_S$, we estimate the $K_S$ magnitudes based on the mean $K_S-W1$ color. For the 1730 host galaxies with both $K_S$ and $W1$ magnitudes, we have an average offset of $-0.43$~mag with a dispersion of $0.04$~mag and a standard error of $0.001$~mag. Similar to the $J$ and $H$ bands, a galaxy detected in neither 2MASS or WISE is given an upper limit of $K_S>15.6$~mag, matching the faintest host magnitude in the total sample.

\begin{figure*}
\begin{minipage}{\textwidth}
\centering
\subfloat{{\includegraphics[width=0.95\textwidth]{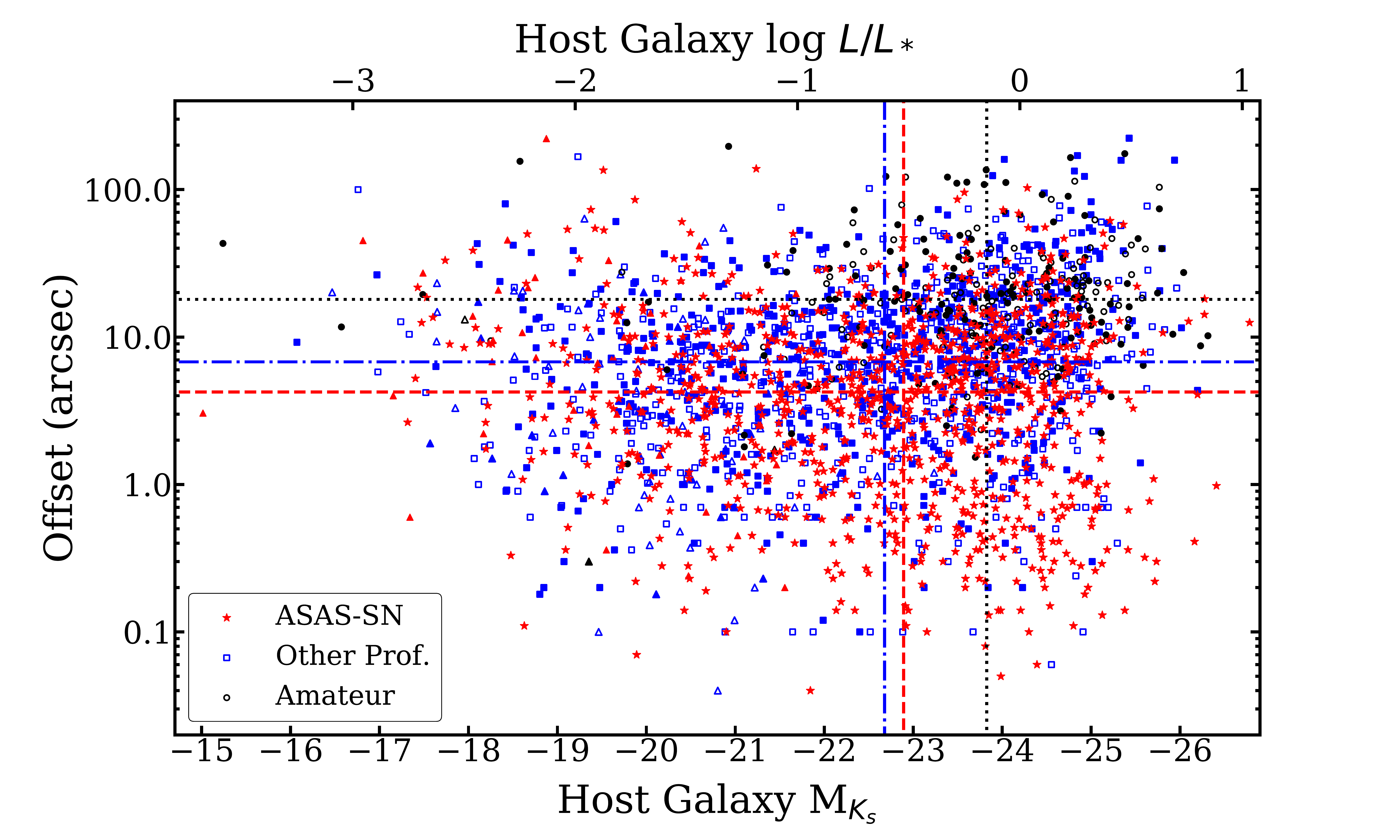}}}

\centering
\subfloat{{\includegraphics[width=0.95\textwidth]{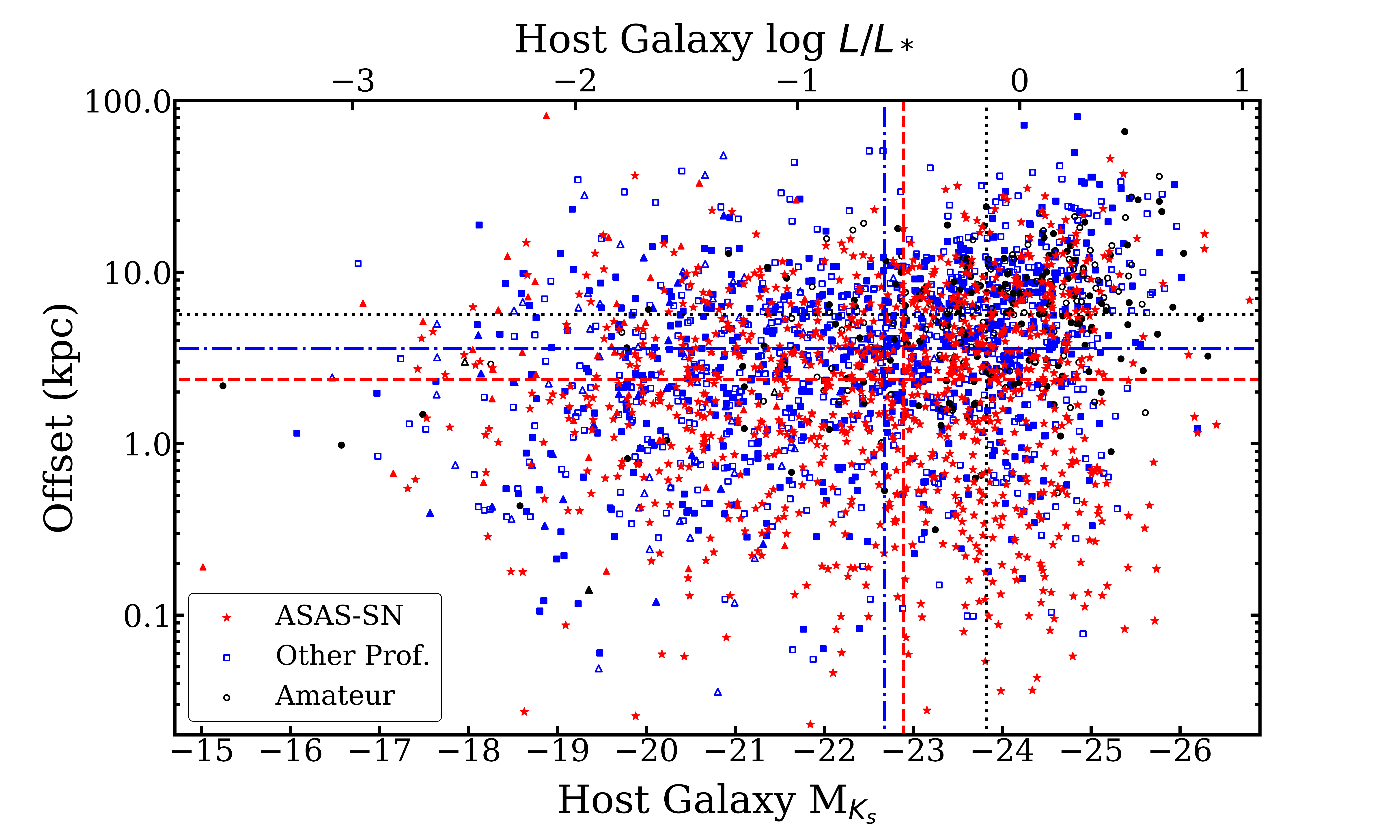}}}
\caption{The host galaxy offsets and absolute magnitudes, $M_{K_S}$, for the full 2014-2020 sample. The offsets are in arcseconds for the \emph{Upper Panel} and kiloparsecs for the \emph{Lower Panel}. The top scale gives $L/L_*$ for the hosts using $M_{\star ,K_S}=-24.2$~mag \citep[][]{kochanek01}. Red stars, blue squares, and black circles are SNe discovered by ASAS-SN, other professional surveys, and amateurs. Filled points represent SNe discovered or recovered by ASAS-SN. Triangles represent hosts without 2MASS or WISE data where the magnitudes are upper limits. Median host magnitudes and offsets are marked by dashed (ASAS-SN), dash-dotted (Other Professionals), and dotted lines (Amateurs) in the colors of their discovery source.}
\label{fig:offmag}
\end{minipage}
\end{figure*}


\begin{figure*}
\begin{minipage}{\textwidth}
\centering
\subfloat{{\includegraphics[width=0.80\textwidth]{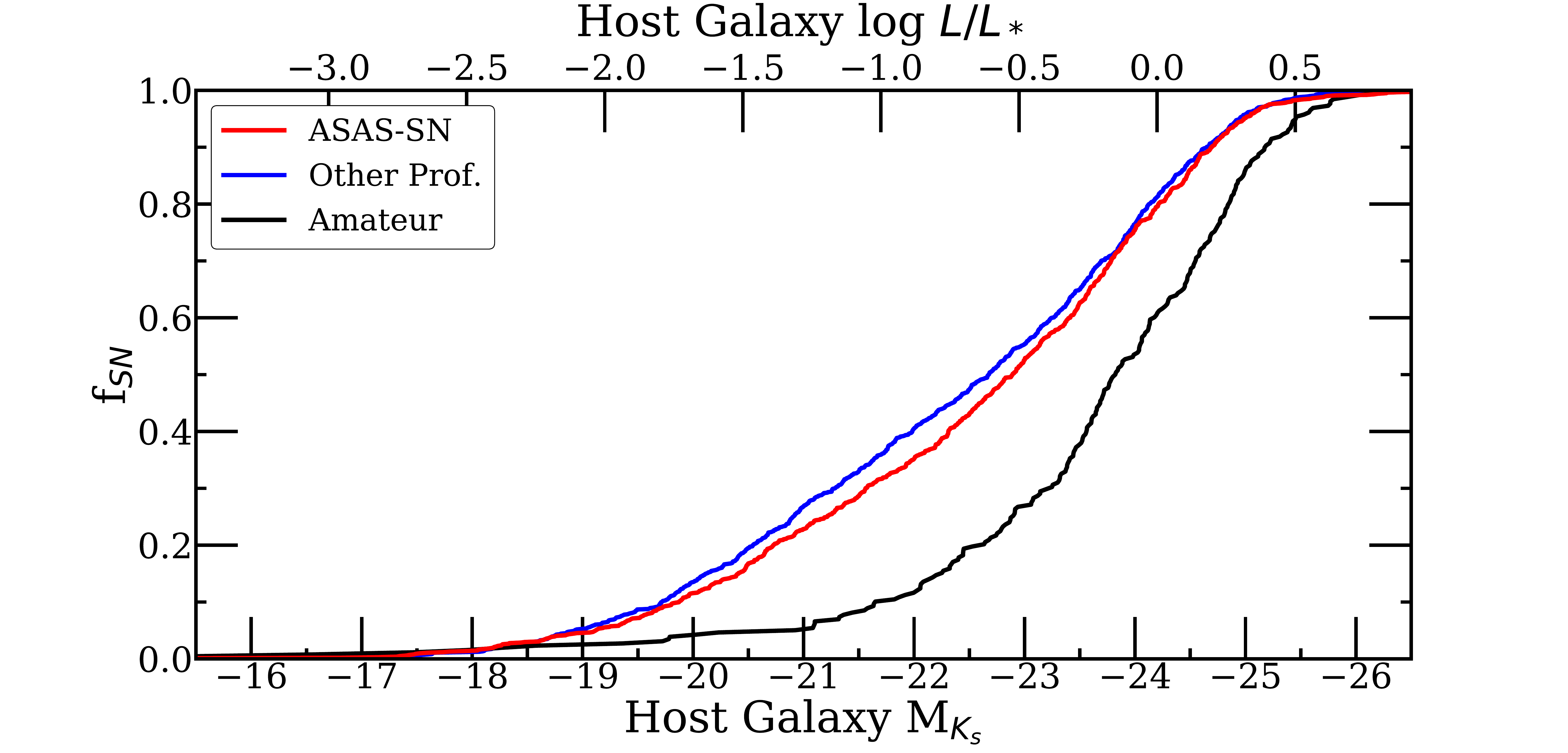}}}

\centering
\subfloat{{\includegraphics[width=0.80\textwidth]{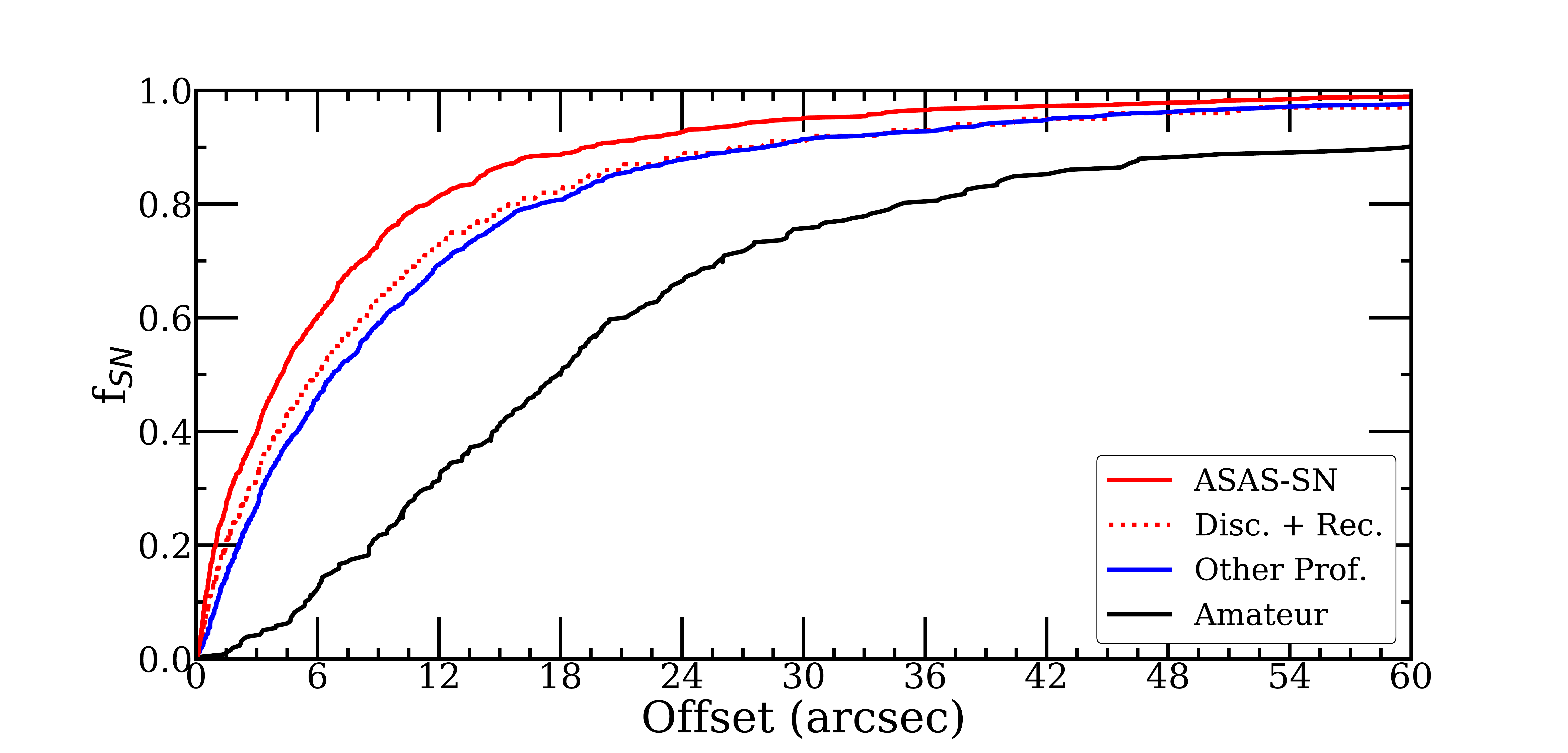}}}

\centering
\subfloat{{\includegraphics[width=0.80\textwidth]{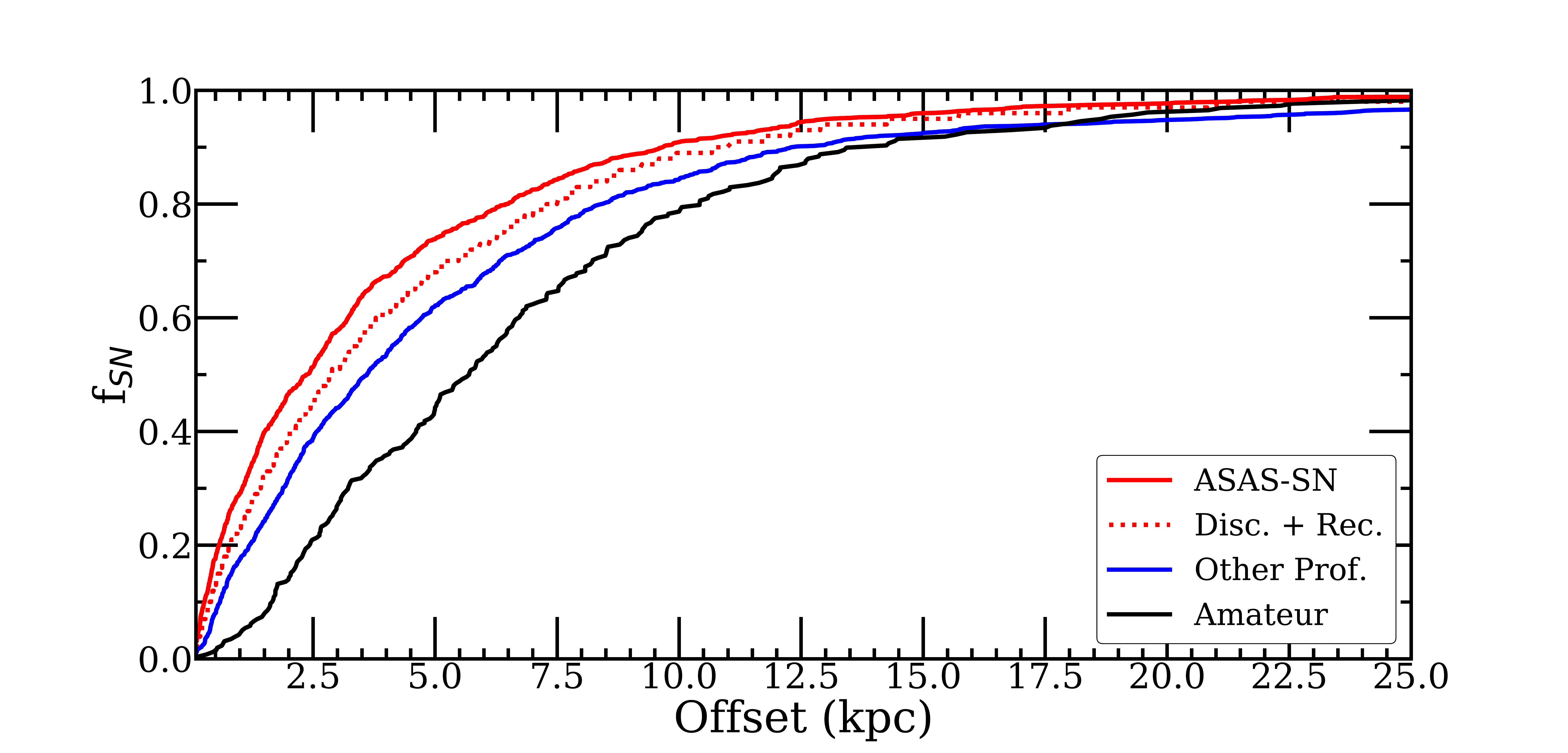}}}
\caption{Normalized cumulative distributions for the total 2014-2020 sample in host galaxy absolute magnitudes, $M_{K_S}$ (\emph{Upper Panel}), supernova offsets in arcseconds (\emph{Center Panel}), and supernova offsets in kiloparsecs (\emph{Bottom Panel}). ASAS-SN discovered SNe are in red, other professional surveys are in blue, and SNe discovered by amateur are in black. The dotted red lines in the offset distribution plots are for ASAS-SN discoveries and recoveries.}

\label{fig:offmag_dist}
\end{minipage}
\end{figure*}


\section{Analysis of the Sample}
\label{sec:analysis}

From 2014 May 01, when ASAS-SN began operating in both hemispheres, to 2020 December 31, 2427 bright SNe were discovered. This total excludes SNe with $m_{peak}>17.0$~mag discovered prior to 2018 and $m_{peak}>18.0$~mag after 2018, as well as unclassified ASAS-SN discoveries. Figure~\ref{fig:histogram} displays the monthly discovery rate of bright SNe from 2012 to 2020. Milestones in the history of ASAS-SN are marked. ASAS-SN was originally built because it appeared that local, bright supernova samples were significantly incomplete. The doubling of the discovery rates between 2012 to 2014 and 2014 to 2018 makes it clear that the problem was real.

After ASAS-SN transitioned to using $g$-band, it made sense to use a limiting magnitude of $g=18$~mag for Figure~\ref{fig:histogram}. While much of the doubling in discovery rate is due to switching to the fainter limit, redoing the rates of the earlier time period with this limit would not lead to a similar doubling prior to 2018. The advent of ZTF and ATLAS in this period resulted in a larger percentage of bright SNe being discovered outside of ASAS-SN. Then in early 2020, ASAS-SN's discovery rate dropped dramatically due to the closures caused by the COVID-19 pandemic. Despite this, ASAS-SN continued to discover or recover about half to two-thirds of bright SNe. The existence of the three partially overlapping surveys ensures that the bright supernova samples are now highly complete.

Over the 2018 to 2020 period, ASAS-SN discovered 443 SNe and recovered 519 other SNe for a total statistical sample of 1706 discovered or recovered ASAS-SN SNe from 2014 May to 2020. Among the external discoveries in the recent period, other professional surveys discovered 952 SNe and amateurs discovered 83 SNe. While ASAS-SN remained the top contributor of bright SNe, ATLAS, ZTF, and Gaia all now surpass amateurs in discoveries of bright SNe where only ASAS-SN and ATLAS did so in 2017 \citep[][]{holoien19a}.

Figure~\ref{fig:piechart} shows the breakdown of supernova types into their basic classes of Type~Ia, Type~II and Type~Ib/Ic where subclasses such as IIb and IIn are included as Type~II to simplify the diagrams. Superluminous SNe (9) are not included nor are 4 incompletely typed SNe (3 Type~I SNe and 1 ``young'' core-collapse supernova) and 142 untyped ASAS-SN SNe. The ratio of untyped to typed SNe was greatest in 2020 due to the closures caused by the COVID-19 pandemic. Compared to the type distribution of all $g<18$~mag SNe, the ASAS-SN discoveries are biased towards Type~Ia SNe and the amateur discoveries are biased towards core-collapse SNe. As we discuss below, amateur searches are biased towards luminous star forming galaxies, leading to a bias towards finding core-collapse SNe. They are effectively all-sky like ASAS-SN if biased in the galaxies observed, and the rapid rise times of core-collapse SNe also reduces the ability of deeper surveys to identify SNe before they approach their peak brightness. The combined distributions of the ASAS-SN and amateur discoveries are very similar to the distribution of all SNe in the new sample. The other professional surveys are more dominated by fainter SNe and do not show the type trade off with the amateurs. The overall distributions are similar to the ``ideal magnitude-limited sample'' of \citet{li11a}, with $79.2^{+4.2}_{-5.5}$\% Type~Ia, $16.6^{+5.0}_{-3.9}$\% Type~II and $4.1^{+1.6}_{-1.2}$\% Type~Ib/Ic. However, there are differences that are likely real. While \citet{li11a} find that a finite observing cadence reduces the fraction of Type~Ia SNe in favor of Type~II SNe, the effect is modest for the cadence of modern surveys. The most noticeable of these differences is the larger fraction of Type~Ib/Ic SNe. The total ASAS-SN sample for 2014 to 2020 includes 1655 Type~Ia SNe, 166 Type~Ib/Ic SNe, 590 Type~II SNe, and 12 superluminous SNe. 


\begin{figure}

\centering
\subfloat{{\includegraphics[width=1\linewidth]{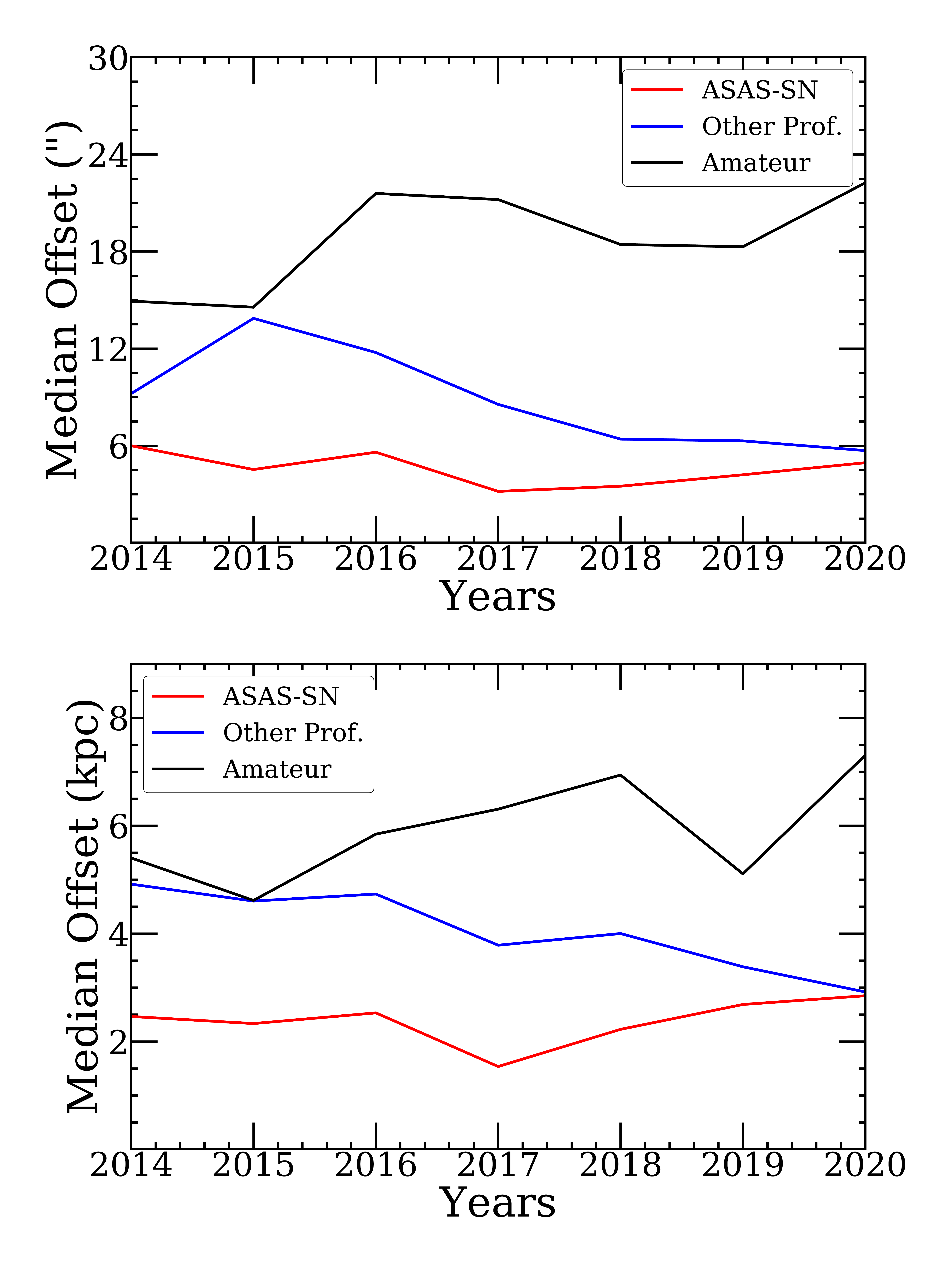}}}

\caption{Annual median host offsets of the supernova. Median supernova host offsets are in arcseconds (\emph{Upper Panel}) and kiloparsecs (\emph{Lower Panel}) where ASAS-SN offsets are in red, other professional survey offsets are in blue, and amateur offsets are in black.}

\label{fig:med_offset}
\end{figure}


\begin{figure}

\centering
\subfloat{{\includegraphics[width=1.00\linewidth]{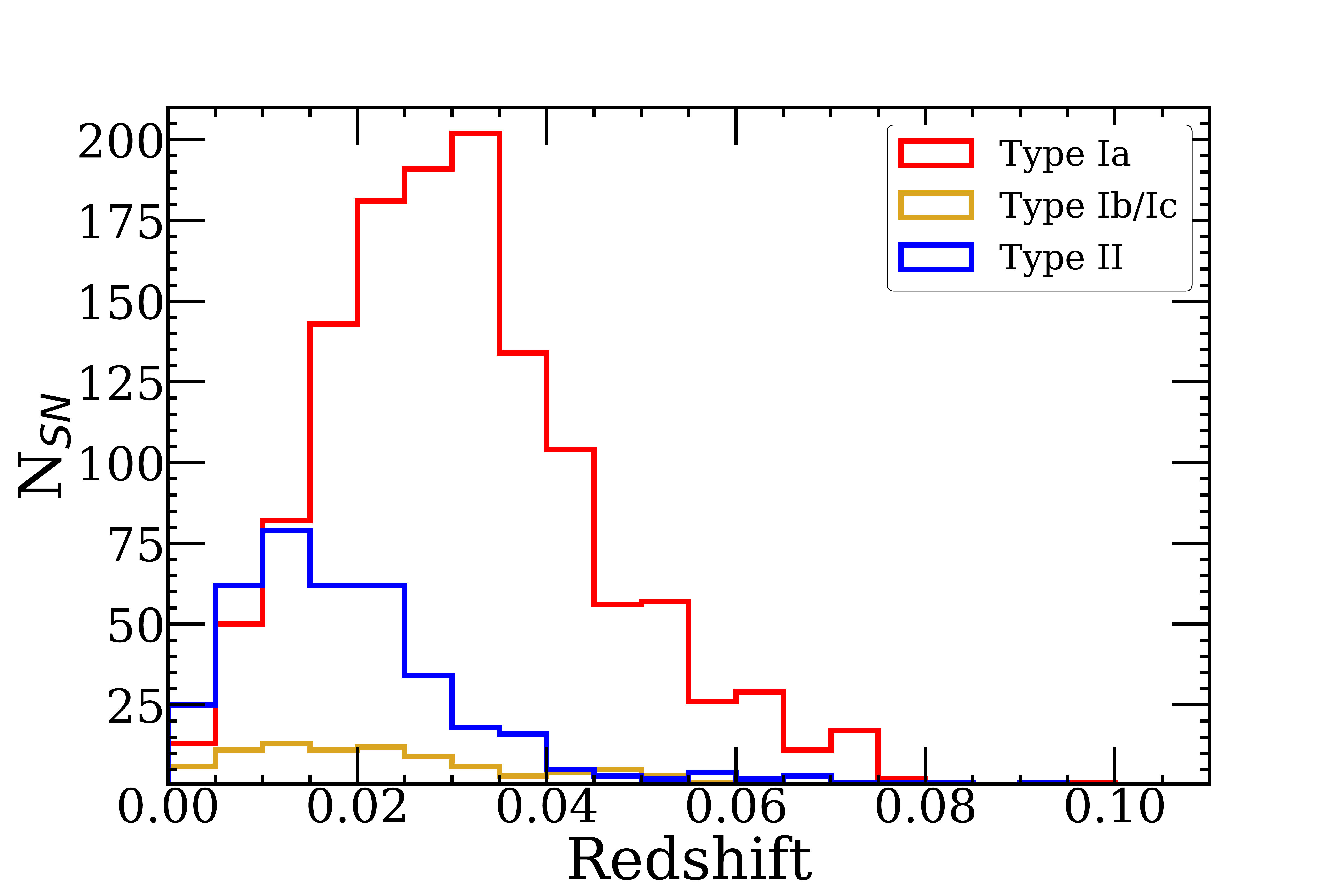}}}
\caption{Redshifts for the total sample of bright SNe discovered or recovered by ASAS-SN from 2014 to 2020 with a bin width of $z=0.005$. Type~Ia SNe are shown in red, Type~Ib/Ic SNe are in dark yellow, and Type~II SNe are in blue with similar sub-class distribution to Figure \ref{fig:piechart}.}
\label{fig:redshift}
\end{figure}


\begin{figure*}

\begin{minipage}{\textwidth}
\centering
\subfloat{{\includegraphics[width=.75\linewidth]{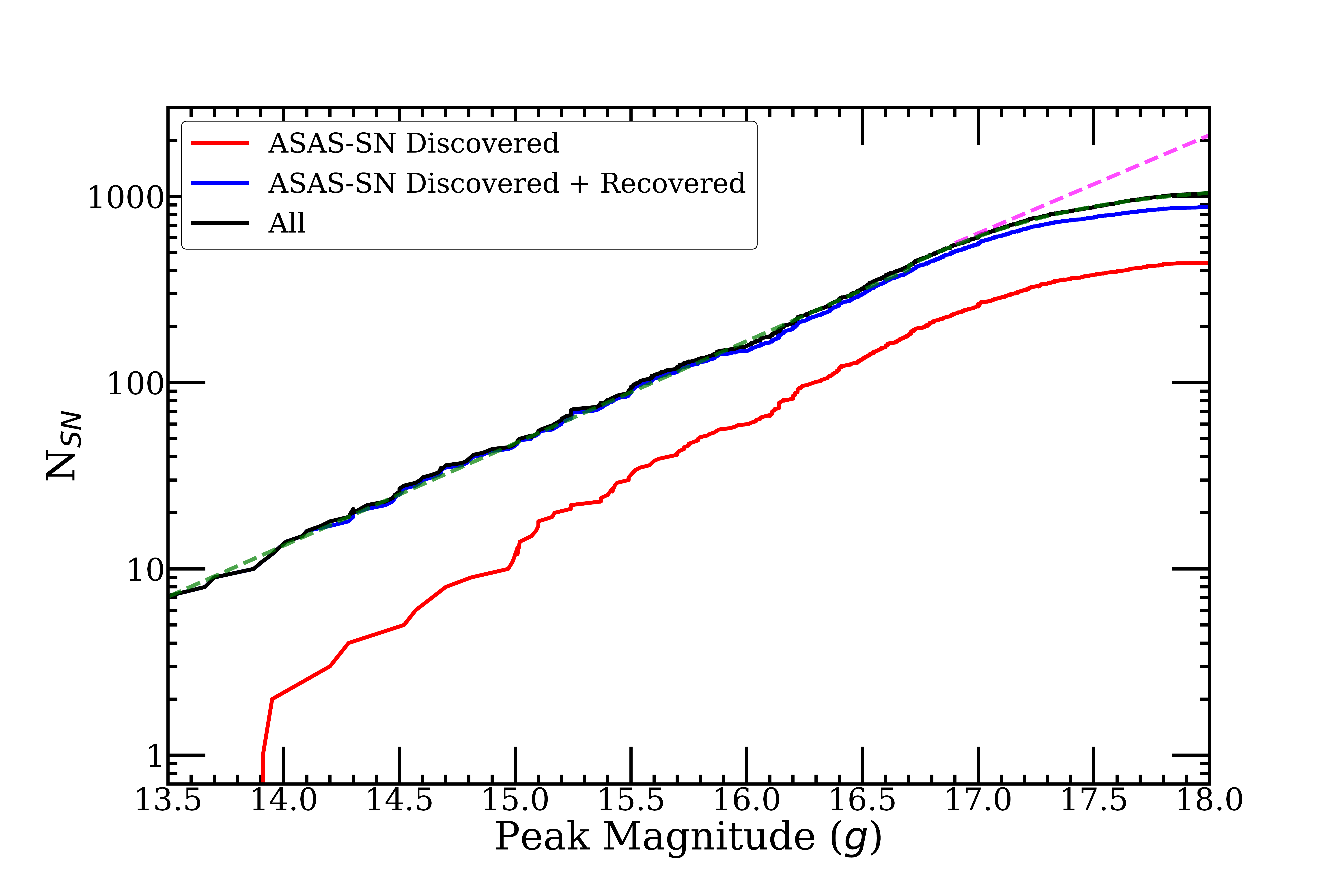}}}
\caption{Cumulative peak $g$-band magnitude distributions of ASAS-SN discoveries (red), discoveries plus recoveries (blue), and all $g\leq18$~mag SNe (black). The dashed green line is the broken power-law fit to the ASAS-SN discovery plus recovery sample (see text). It is mostly invisible under the black line. The dashed magenta line is the Euclidean distribution.}

\label{fig:mag_dist}
\end{minipage}
\end{figure*}

Figures~\ref{fig:offmag} and \ref{fig:offmag_dist} show the distribution of the absolute host magnitude, $M_{K_S}$, and the supernova offsets in arcseconds and kiloparsecs. The upper luminosity scale gives $L/L_*$ for $M_{*,K_S}=-24.2$~mag \citep[][]{kochanek01}. As found in \citet{holoien19a}, amateurs tend to discover SNe at larger offsets and in more luminous galaxies than ASAS-SN or the other professional surveys. For a given observing effort, focusing on luminous star forming galaxies will have the highest yield of SNe and will create the bias of the amateurs towards finding core-collapse SNe. While the host luminosity distributions of ASAS-SN and other professional surveys are essentially identical, ASAS-SN continues to find smaller median offsets in both angular (3\farcs{9} vs 6\farcs{1}) and physical (2.5 kpc vs 3.4 kpc) units for 2018 to 2020. The $7\farcs0$ pixel scale of ASAS-SN's telescopes causes many possible SNe to appear within the nuclear regions of their hosts, so this region could not be avoided in ASAS-SN. While ASAS-SN is better at discovering SNe close to their host nucleus, ASAS-SN recoveries do not follow this pattern due to being discovered by others. Figure~\ref{fig:med_offset} shows the median offsets of SNe for each discovery source per year. While ASAS-SN and amateur median offsets remain relatively consistent, the other professional surveys have steadily smaller median offsets, and their median offsets nearly equal ASAS-SN's in 2020. \citet{fremling20} find similar results in their comparison of ZTF and ASAS-SN (see their Figure 7).

Of the 971 ASAS-SN SNe, 28\% (272) came from hosts without reported redshifts and 2\% (19) came from uncataloged hosts or were hostless. Of the 1457 Non-ASAS-SN SNe, 23\% (335) came from hosts without measured redshifts and $<$1\% (9) came from uncataloged hosts or were hostless. All SNe in the 2018-2020 sample came from a cataloged host galaxy. The distribution of ASAS-SN discovered or recovered supernova redshift by type in Figure~\ref{fig:redshift} follows the expected trends where the more luminous Type~Ia SNe are detected at higher redshifts than core-collapse SNe. The median for the Type~Ia SNe distribution is $z=0.030$ while the median for the Type~II SNe redshift is $z=0.020$, and the median redshift for Type~Ib/Ic SNe is $z=0.017$.

Figure~\ref{fig:mag_dist} shows the cumulative histogram of the peak magnitudes of SNe discovered by ASAS-SN, SNe discovered or recovered by ASAS-SN, and all SNe with $g\leq18$~mag and peak magnitudes from the ASAS-SN $g$-band light curves. This excludes SNe where we used the brightest observed magnitude for their peak magnitude. Very bright SNe ($m_{peak}\lesssim14.5$~mag; \citealt{holoien19a}) are most commonly discovered by amateurs or the Distance Less Than 40 Mpc (DLT40; \citealt{tartaglia18}) survey\footnote{\url{http://dark.physics.ucdavis.edu/dlt40/DLT40}}. In 2018 to 2020, ASAS-SN recovered or discovered all SNe of $m_{peak}\lesssim15$~mag, excluding supernova DLT18aq (SN~2018ivc). This supernova was too close to the nucleus of the AGN~NGC1068 for ASAS-SN's resolution to differentiate the two. The supernova was confused with the variable AGN in ASAS-SN's data, so it was ignored.

We modeled the unbinned differential distribution of $g\leq18$~mag SNe as a broken power law with Markov Chain Monte Carlo (MCMC) methods, holding the bright SNe slope fixed to the Euclidean value of 0.6. We find a break magnitude of $m_{break}=16.74\pm0.04$~mag and a faint slope of $-0.66\pm0.08$. This is significantly better than in the $V$-band sample where $m_{break}=16.24\pm0.11$~mag \citep{holoien19a}. 

The completeness drops rapidly for $g\gtrsim17$~mag. The integral completeness of the sample is $100\%$ at $g=16.5$~mag, $90\pm2\%$ at $g=17.0$~mag, $57\pm2\%$ at $g=17.5$~mag, and $31\pm1\%$ at $g=18.0$~mag relative to a Euclidean model. The differential completeness is $100\%$ at $g=16.5$~mag, $47\pm4\%$~mag at $g=17.0$~mag, $11\pm1\%$ at $g=17.5$~mag, and $2.6\pm0.4\%$ at $g=18.0$~mag. The Euclidean model modestly underestimates the completeness by not using a full cosmological model, time dilation, and K-corrections.


\section{Conclusions}
\label{sec:disc}

We catalog the 1478 bright SNe discovered, recovered, or missed by ASAS-SN from 2018 to 2020. The complete ASAS-SN sample starting from 2014 now totals 2427 bright SNe with 971 SNe discovered by ASAS-SN and 735 SNe independently recovered by ASAS-SN. With the start of using $g$-band filters in 2017 and the complete conversion to $g$-band in 2018, ASAS-SN's discovery rate significantly increased. As the ATLAS and ZTF discovery rates increased, the ASAS-SN discovery rate declined, and there was a large drop due to COVID-19 in 2020. Amateur discoveries have steadily dropped. They discovered 34 SNe in 2017 but only about 28 SNe per year from 2018 to 2020. ASAS-SN still discovers SNe closer to the nuclei of their host galaxy than the other professional surveys or amateurs (Figure~\ref{fig:offmag}), but other professional surveys appear to have closed the gap by 2020. Along with offsets, the all-sky nature of ASAS-SN allows for a complete view of the sky every night improving the completeness of the data beyond any other survey during this time period. With the increase to a limiting magnitude of $g\leq18$~mag, our sample is complete up to a peak magnitude $m_{peak}=16.7$~mag, 90\% complete for $m_{peak}\leq17.0$~mag, and 30\% complete for $m_{peak}\leq18.0$~mag. This is a significant increase from the previous $V$-band catalogs where our sample was only complete up to $m_{peak}=16.2$~mag and only 70\% complete for SNe brighter than $m_{peak}\leq17.0$~mag \citep[][]{holoien19a}. 

The primary purpose of the ASAS-SN catalogs is to enable statistical studies. Until recently, the largest, local statistical sample of SNe came from the 137 SNe studied in \citet{cappellaro99}, the SDSS survey (72 Type~Ia SNe at $z<0.15$, \citealt{dilday10}), the Lick Observatory Supernova Search (LOSS; 74 Type~Ia and 106 core-collapse SNe in \citealt{li11a} and 274 Type~Ia and 440 core-collapse SNe in \citealt{li11b}), and 90 Type~Ia SNe from the Palomar Transient Factory \citep{frohmaier19}. The local statistical sample was greatly expanded to 875 Type~Ia and 373 core-collapse SNe from ZTF in \citet{fremling20} and \citet{perley20} although their completeness corrections are only approximate. A volume limited subset of 298 ZTF Type~Ia SNe from this sample were analyzed by \citet{sharon22}. ASAS-SN has slowly been building towards such statistical analyses, starting with an analysis of the Type~Ia supernova rate as a function of stellar mass in \citet{brown19} using 476 Type~Ia SNe, and estimates of the volumetric Type~Ia supernova rate including luminosity functions for major subtypes of Type~Ia SNe in \citet{chen22} using 247 Type~Ia SNe and in Desai et al. (\textit{in prep.}) using 400 Type~Ia SNe. With this extension to the ASAS-SN catalogs to a statistical sample of 1655 Type~Ia SNe and 756 core-collapse SNe, the objective is to use these samples to do expanded analyses of the Type~Ia SNe (rates and luminosity functions of host properties) and to carry out similar analyses for the core-collapse supernova sample.


\section*{Acknowledgments}

The authors thank Las Cumbres Observatory and its staff for their continued support of ASAS-SN.  

CSK and KZS are supported by NSF grants AST-1814440 and AST-1908570. Support for Support for T.W.-S.H. was provided by NASA through the NASA Hubble Fellowship grant HST-HF2-51458.001-A awarded by the Space Telescope Science Institute (STScI), which is operated by the Association of Universities for Research in Astronomy, Inc., for NASA, under contract NAS5-26555. Support for JLP is provided in part by ANID through the Fondecyt regular grant 1191038 and through the Millennium Science Initiative grant ICN12-009, awarded to The Millennium Institute of Astrophysics, MAS. JFB was supported by NSF Grant No. PHY-2012955. M.D.S. was funded in part by an Experiment grant (\# 28021) from the Villum FONDEN, and by a project 1 grant (\#8021-00170B) from the Independent Research Fund Denmark (IRFD).

ASAS-SN is funded in part by the Gordon and Betty Moore Foundation through grants GBMF5490 and GBMF10501 to the Ohio State University, the Mt. Cuba Astronomical Foundation, the Center for Cosmology and AstroParticle Physics (CCAPP) at OSU, the Chinese Academy of Sciences South America Center for Astronomy (CAS-SACA), and the Villum Fonden (Denmark). Development of ASAS-SN has been supported by NSF grant AST-0908816, the Center for Cosmology and Astroparticle Physics, Ohio State University, the Mt. Cuba Astronomical Foundation, and by George Skestos. This work is based on observations made by ASAS-SN. We wish to extend our special thanks to those of Hawaiian ancestry on whose sacred mountains of Maunakea and Haleakal\={a}, we are privileged to be guests. Without their generous hospitality, the observations presented herein would not have been possible.

Software used: ASTROPY (\citealt{astropy22}), {\sc Iraf} \citep[][]{tody86}, NUMPY (\citealt{harris20}), Matplotlib (\citealt{hunter07}).

Facilities: Laser Interferometer Gravitational-Wave Observatory (USA), Virgo (Italy), Haleakala Observatories (USA), Cerro Tololo International Observatory (Chile), McDonald Observatory (USA), South African Astrophysical Observatory (South Africa).

This research uses data obtained through the Telescope Access Program (TAP), which has been funded by ``the Strategic Priority Research Program-The Emergence of Cosmological Structures'' of the Chinese Academy of Sciences (Grant No.11 XDB09000000) and the Special Fund for Astronomy from the Ministry of Finance.

This research has made use of the XRT Data Analysis Software (XRTDAS) developed under the responsibility of the ASI Science Data Center (ASDC), Italy. At Penn State the NASA {\swift} program is support through contract NAS5-00136.

This research was made possible through the use of the AAVSO Photometric All-Sky Survey (APASS), funded by the Robert Martin Ayers Sciences Fund.

This research has made use of data provided by Astrometry.net \citep{barron08,lang10}.

This paper uses data products produced by the OIR Telescope Data Center, supported by the Smithsonian Astrophysical Observatory.

Observations made with the NASA Galaxy Evolution Explorer (GALEX) were used in the analyses presented in this manuscript. Some of the data presented in this paper were obtained from the Mikulski Archive for Space Telescopes (MAST). STScI is operated by the Association of Universities for Research in Astronomy, Inc., under NASA contract NAS5-26555. Support for MAST for non-HST data is provided by the NASA Office of Space Science via grant NNX13AC07G and by other grants and contracts.

Funding for the Sloan Digital Sky Survey IV has been provided by the Alfred P. Sloan Foundation, the U.S. Department of Energy Office of Science, and the Participating Institutions. SDSS acknowledges support and resources from the Center for High-Performance Computing at the University of Utah. The SDSS web site is www.sdss.org.

SDSS is managed by the Astrophysical Research Consortium for the Participating Institutions of the SDSS Collaboration including the Brazilian Participation Group, the Carnegie Institution for Science, Carnegie Mellon University, Center for Astrophysics | Harvard \& Smithsonian (CfA), the Chilean Participation Group, the French Participation Group, Instituto de Astrofísica de Canarias, The Johns Hopkins University, Kavli Institute for the Physics and Mathematics of the Universe (IPMU) / University of Tokyo, the Korean Participation Group, Lawrence Berkeley National Laboratory, Leibniz Institut für Astrophysik Potsdam (AIP), Max-Planck-Institut für Astronomie (MPIA Heidelberg), Max-Planck-Institut f\"{u}r Astrophysik (MPA Garching), Max-Planck-Institut f\"{u}r Extraterrestrische Physik (MPE), National Astronomical Observatories of China, New Mexico State University, New York University, University of Notre Dame, Observatório Nacional / MCTI, The Ohio State University, Pennsylvania State University, Shanghai Astronomical Observatory, United Kingdom Participation Group, Universidad Nacional Autónoma de México, University of Arizona, University of Colorado Boulder, University of Oxford, University of Portsmouth, University of Utah, University of Virginia, University of Washington, University of Wisconsin, Vanderbilt University, and Yale University.

The Pan-STARRS1 Surveys (PS1) and the PS1 public science archive have been made possible through contributions by the Institute for Astronomy, the University of Hawaii, the Pan-STARRS Project Office, the Max-Planck Society and its participating institutes, the Max Planck Institute for Astronomy, Heidelberg and the Max Planck Institute for Extraterrestrial Physics, Garching, The Johns Hopkins University, Durham University, the University of Edinburgh, the Queen's University Belfast, the Harvard-Smithsonian Center for Astrophysics, the Las Cumbres Observatory Global Telescope Network Incorporated, the National Central University of Taiwan, the Space Telescope Science Institute, the National Aeronautics and Space Administration under Grant No. NNX08AR22G issued through the Planetary Science Division of the NASA Science Mission Directorate, the National Science Foundation Grant No. AST-1238877, the University of Maryland, Eotvos Lorand University (ELTE), the Los Alamos National Laboratory, and the Gordon and Betty Moore Foundation.

This publication makes use of data products from the Two Micron All Sky Survey, which is a joint project of the University of Massachusetts and the Infrared Processing and Analysis Center/California Institute of Technology, funded by NASA and the National Science Foundation.

This publication makes use of data products from the Wide-field Infrared Survey Explorer, which is a joint project of the University of California, Los Angeles, and the Jet Propulsion Laboratory/California Institute of Technology, funded by NASA.

This research is based in part on observations obtained at the Southern Astrophysical Research (SOAR) telescope, which is a joint project of the Minist\'{e}rio da Ci\^{e}ncia, Tecnologia, e Inova\c{c}\~{a}o (MCTI) da Rep\'{u}blica Federativa do Brasil, the U.S. National Optical Astronomy Observatory (NOAO), the University of North Carolina at Chapel Hill (UNC), and Michigan State University (MSU).

The Liverpool Telescope is operated on the island of La Palma by Liverpool John Moores University in the Spanish Observatorio del Roque de los Muchachos of the Instituto de Astrofisica de Canarias with financial support from the UK Science and Technology Facilities Council.

This research has made use of the NASA/IPAC Extragalactic Database (NED),
which is operated by the Jet Propulsion Laboratory, California Institute of Technology,
under contract with the National Aeronautics and Space Administration.

\section*{Data Availability}
All data analyzed and discussed in this catalog is available in a machine-readable form in the online journal as supplementary material. A portion is shown in Tables~\ref{table:asassn_sne}, \ref{table:other_sne}, \ref{table:asassn_hosts}, and \ref{table:other_hosts} for guidance regarding its form and content.

\bibliographystyle{mnras2}
\bibliography{bibliography_catalogs_new,bibliography_SNe_new}
\fontsize{8}{10.2}\selectfont

\section*{Affiliations}
\textit{
  $^{1}$ Department of Astronomy, The Ohio State University, 140 West 18th Avenue, Columbus, OH 43210, USA \\
  $^{2}$ The Observatories of the Carnegie Institution for Science, 813 Santa Barbara Street, Pasadena, CA 91101, USA \\
  $^{3}$ Center for Cosmology and AstroParticle Physics (CCAPP), The Ohio State University, 191 W. Woodruff Ave., Columbus, OH 43210, USA \\
  $^{4}$ Institute for Astronomy, University of Hawai'i, 2680 Woodlawn Drive, Honolulu, HI 96822, USA \\
  $^{5}$ N\'ucleo de Astronom\'ia de la Facultad de Ingenier\'ia y Ciencias, Universidad Diego Portales, Av. Ej\'ercito 441, Santiago, Chile \\
  $^{6}$ MAS, Millennium Institute of Astrophysics, Santiago, Chile \\
  $^{7}$ European Southern Observatory, Alonso de C\'ordova 3107, Casilla 19, Santiago, Chile \\
  $^{8}$ Department of Astronomy, University of California Berkeley, Berkeley CA 94720, USA \\
  $^{9}$ Coral Towers Observatory, Cairns, Queensland 4870, Australia \\
  $^{10}$ Astrophysics Research Institute, Liverpool John Moores University, 146 Brownlow Hill, Liverpool L3 5RF, UK \\
  $^{11}$ Center for Data Intensive and Time Domain Astronomy, Department of Physics and Astronomy, Michigan State University, East Lansing, MI 48824, USA \\
  $^{12}$ Department of Physics, The Ohio State University, 191 W. Woodruff Ave., Columbus, OH 43210, USA \\
  $^{13}$ Kavli Institute for Astronomy and Astrophysics, Peking University, Yi He Yuan Road 5, Hai Dian District, Beijing 100871, China \\
  $^{14}$ Department of Astronomy and Astrophysics, University of California, Santa Cruz, CA 92064, USA \\
  $^{15}$ Department of Particle Physics and Astrophysics, Weizmann Institute of Science, 234 Herzl St, 7610001 Rehovot, Israel \\
  $^{16}$ Instituto de Astrofísica, Pontificia Universidad Catolica de Chile, Instituto Milenio de Astrofisica (MAS), Vicuna Mackenna 4860, Macul, Santiago, Chile\\
  $^{17}$ Harvard-Smithsonian Center for Astrophysics, 60 Garden St., Cambridge, MA 02138, USA \\
  $^{18}$ Department of Physics and Astronomy, Aarhus University, Ny Munkegade 120, DK-8000 Aarhus C, Denmark \\
  $^{19}$ Las Campanas Observatory, Carnegie Observatories, Casilla 601, La Serena, Chile \\
  $^{20}$ Department of Physics and Astronomy, Michigan State University, East Lansing, MI 48824, USA \\
  $^{21}$ Sternberg Astronomical Institute, Moscow State University, Universitetskii pr. 13, 119992 Moscow, Russia \\
  $^{22}$ National Research Council Research Associate, National Academy of Sciences, Washington, DC 20001, USA, resident at Naval Research Laboratory, Washington, DC 20375, USA \\
  $^{23}$ Physics \& Astronomy Department, Georgia State University, 25 Park Place, Atlanta, GA 30302, USA \\
  $^{24}$ Rochester Academy of Science, 1194 West Avenue, Hilton, NY 14468, USA \\
  $^{25}$ Runaway Bay Observatory, 1 Lee Road, Runaway Bay, Queensland 4216, Australia \\
  $^{26}$ Samford Valley Observatory (Q79), Queensland, 4520, Australia \\
  $^{27}$ DogsHeaven Observatory, SMPW Q25 CJ1 LT10, Brasilia, DF 71745-501, Brazil \\
  $^{28}$ Association Francaise des Observateurs d'Etoiles Variables (AFOEV), Observatoire de Strasbourg, 11 Rue de l'Universite, France \\
  $^{29}$ Moondyne Observatory, 61 Moondyne Rd, Mokine, WA 6401, Australia \\
  $^{30}$ Slate Ridge Observatory, 1971 Haverton Drive, Reynoldsburg, OH 43068, USA \\
  $^{31}$ Observatorio Uraniborg, MPC Z55, Ecija, Sevilla, Spain \\
  $^{32}$ Observadores de Supernovas (ObSN), Spain \\
  $^{33}$ Observatory Inmaculada del Molino, Sevilla, Spain \\
  $^{34}$ Al Sadeem Observatory, Al Wathba South, Abu Dhabi, UAE \\
  $^{35}$ Observatorio Cerro del Viento-MPC I84, Badajoz, Spain \\
  $^{36}$ AAVSO, 185 Alewife Brook Parkway, Suite 410, Cambridge, MA 02138, USA \\
  $^{37}$ Prince George Astronomical Observatory, 7365 Tedford, Prince George, BC V2N 6S2, Canada \\
  $^{38}$ Variable Star Observers League in Japan, 7-1 Kitahatsutomi, Kamagaya, Chiba 273-0126, Japan \\
  $^{39}$ Antelope Hills Observatory, 980 Antelope Drive West, Bennett, CO 80102, USA \\
  $^{40}$ Roof Observatory Kaufering, Lessingstr. 16, D-86916 Kaufering, Germany \\
  $^{41}$ Peter Observatory, unit 316 Sandcastles 392 Marine Pde, Labrador, Qld 4215, Australia \\
  $^{42}$ Virtual Telescope Project, Via Madonna de Loco, 47-03023 Ceccano (FR), Italy \\
  $^{43}$ Kleinkaroo Observatory, Calitzdorp, St. Helena 1B, P.O. Box 281, 6660 Calitzdorp, Western Cape, South Africa \\
  $^{44}$ Departamento de Astronom\'{\i}a y Astrof\'{\i}sica, Universidad de Valencia, E-46100 Burjassot, Valencia, Spain \\
  $^{45}$ Observatorio Astron\'omico, Universidad de Valencia, E-46980 Paterna, Valencia, Spain \\
  $^{46}$ Mount Vernon Observatory, 6 Mount Vernon Place, Nelson, New Zealand \\
  $^{47}$ Post Observatory, Lexington, MA 02421, USA \\
  $^{48}$ Alexander Observatory, Brisbane, Australia \\
  $^{49}$ First Light Observatory Systems, 17908 NE 391st St, Amboy, WA 98601, USA \\
  $^{50}$ INAF – Osservatorio Astronomico di Padova, Vicolo dell’Osservatorio 5, I-35122 Padova, Italy \\
  $^{51}$ Brisbane Girls Grammar School - Dorothy Hill Observatory, Gregory Terrace, Spring Hill, Queensland 4000, Australia \\
  $^{52}$ Department of Earth and Environmental Sciences, University of Minnesota, 230 Heller Hall, 1114 Kirby Drive, Duluth, MN. 55812, USA \\
  }
\newpage

\begin{landscape}
\begin{table}
\begin{minipage}{\textwidth}
\centering
\fontsize{7}{8}\selectfont
\caption{ASAS-SN Supernovae}
\label{table:asassn_sne}
\begin{tabular}{@{}l@{\hspace{0.15cm}}l@{\hspace{0.15cm}}c@{\hspace{0.15cm}}c@{\hspace{0.15cm}}c@{\hspace{0.15cm}}l@{\hspace{0.15cm}}c@{\hspace{0.15cm}}c@{\hspace{0.15cm}}c@{\hspace{0.15cm}}c@{\hspace{0.15cm}}c@{\hspace{0.15cm}}c@{\hspace{0.15cm}}l@{\hspace{0.15cm}}l@{\hspace{0.15cm}}l@{\hspace{-0.05cm}}} 
\hline
\vspace{-0.14cm}
 & & & & & & & & & & & & & & \\
 & IAU & Discovery & & & & & & & Offset & & Age & & & \\
SN Name & Name & Date & RA$^a$ & Dec.$^a$ & Redshift & $m_{disc}^b$ & $V_{peak}^{c,d}$ & $g_{peak}^{c,d}$ & (arcsec)$^e$ & Type & at Disc.$^f$ & Host Name$^g$ & Discovery Report & Classification Report \\
\vspace{-0.23cm} \\
\hline
\vspace{-0.17cm}
 & & & & & & & & & & & & & &\\
ASASSN-18ae  &  SN 2018br  &  2018-01-06.06  &  03:07:52.53  &  -45:44:22.74  &  0.06281  &  18.0  &  ---  &  17.5  &  7.52  &  Ia  &  ---  &  2MASX J03075327  & \citet{2018TNSTR..30....1S} & \citet{2018TNSCR..82....1B} \\  
ASASSN-18aa  &  SN 2018bg  &  2018-01-07.01  &  05:11:47.86  &  -40:11:43.30  &  0.03000  &  16.5  &  ---  &  16.5  &  0.84  &  Ia  &  0  &  GALEXMSC J051147.86  & \citet{2018TNSTR..21....1S} & \citet{2018TNSCR..48....1B} \\  
ASASSN-18ac  &  SN 2018bq  &  2018-01-08.38  &  11:05:59.60  &  -12:31:37.70  &  0.02563  &  16.2  &  ---  &  16.0  &  0.57  &  Ia  &  ---  &  LCRS B110329.3  & \citet{2018TNSTR..28....1S} & \citet{2018TNSCR..38....1L} \\  
ASASSN-18af  &  SN 2018bs  &  2018-01-09.09  &  03:23:20.73  &  -22:07:00.66  &  0.07000  &  18.0  &  ---  &  17.7*  &  6.23  &  Ia  &  -8  &  2MASX J03232113  & \citet{2018TNSTR..31....1S} & \citet{2018TNSCR..62....1R} \\  
ASASSN-18ag  &  SN 2018ds  &  2018-01-09.46  &  14:48:53.57  &  +38:46:03.68  &  0.03166  &  16.9  &  ---  &  17.2  &  4.17  &  Ia  &  -2  &  MCG +07-30-065  & \citet{2018TNSTR..31....1S} & \citet{2018TNSCR.540....1L} \\  
ASASSN-18aj  &  SN 2018dx  &  2018-01-10.21  &  09:16:12.28  &  +39:03:42.77  &  0.06000  &  17.3  &  ---  &  17.3  &  1.06  &  Ia  &  -1  &  SDSS J091612.25  & \citet{2018TNSTR..34....1B} & \citet{2018TNSCR..47....1L} \\  
ASASSN-18al  &  SN 2018ep  &  2018-01-12.39  &  11:22:40.75  &  +12:01:31.66  &  0.03984  &  17.1  &  ---  &  16.8*  &  3.16  &  Ia  &  2  &  IC 2777  & \citet{2018TNSTR..43....1S} & \citet{2018TNSCR..62....1R} \\  
ASASSN-18an  &  SN 2018gl  &  2018-01-13.57  &  09:58:06.11  &  +10:21:33.62  &  0.01791  &  16.8  &  ---  &  16.1  &  12.50  &  Ia  &  -6  &  NGC 3070  & \citet{2018TNSTR..54....1B} & \citet{2018TNSCR..74....1D} \\  
ASASSN-18am  &  SN 2018gk  &  2018-01-13.64  &  16:35:53.90  &  +40:01:58.30  &  0.03101  &  16.6  &  ---  &  16.1  &  8.55  &  II  &  -3  &  WISE J163554.27  & \citet{2018TNSTR..54....1B} & \citet{2018TNSCR..81....1F} \\  
ASASSN-18ap  &  SN 2018gn  &  2018-01-14.12  &  01:46:42.38  &  +32:30:27.18  &  0.03750  &  17.7  &  ---  &  16.7*  &  1.79  &  II  &  ---  &  KUG 0143+322  & \citet{2018TNSTR..58....1K} & \citet{2018TNSCR..81....1F} \\  
ASASSN-18ao  &  AT 2018gm  &  2018-01-14.32  &  10:21:19.14  &  +20:54:37.33  &  0.04104  &  18.1  &  ---  &  17.7  &  0.77  &  Ia  &  -2  &  SDSS J102119.17  & \citet{2018TNSTR..59....1S} & \citet{2018ATel11180....1F} \\  
ASASSN-18ar  &  SN 2018hv  &  2018-01-15.18  &  03:30:11.15  &  -13:31:22.84  &  0.04098  &  18.3  &  ---  &  17.4  &  3.00  &  Ia  &  -2  &  2MASX J03301134  & \citet{2018TNSTR..76....1F} & \citet{2018TNSCR..81....1F} \\  
ASASSN-18ba  &  SN 2018jm  &  2018-01-20.05  &  05:08:11.96  &  -54:38:41.21  &  0.06400  &  18.1  &  ---  &  17.6  &  1.50  &  Ia  &  0  &  GALEXASC J050811.97  & \citet{2018TNSTR.100....1C} & \citet{2018TNSCR.161....1C} \\  
ASASSN-18az  &  SN 2018jh  &  2018-01-21.43  &  14:21:17.50  &  -06:37:38.78  &  0.02720  &  16.8  &  ---  &  16.5  &  6.05  &  Ia  &  -10  &  2MASX J14211713  & \citet{2018TNSTR..95....1P} & \citet{2018TNSCR.118....1N} \\  
ASASSN-18bj  &  SN 2018kq  &  2018-01-24.34  &  09:24:57.27  &  +40:23:56.15  &  0.02780  &  17.7  &  ---  &  17.4  &  7.45  &  Ic-BL  &  13  &  KUG 0921+406  & \citet{2018TNSTR.111....1K} & \citet{2018TNSCR.120....1N} \\  

\vspace{-0.22cm}
 & & & & & & & & & & & & & &\\
\hline
\end{tabular}
\smallskip
\\
\raggedright
\noindent This table is available in its entirety in a machine-readable form in the online journal. A portion is shown here for guidance regarding its form and content.\\
$^a$ Right ascension and declination are given in the J2000 epoch. \\
$^b$ Discovery magnitudes are $V$- or $g$-band magnitudes from ASAS-SN, depending on the telescope used for discovery. \\
$^c$ Peak $V$- and $g$-band magnitudes are measured from ASAS-SN data. \\
$^d$ Magnitudes marked with a ``*'' are derived from maximum detected magnitude rather than a fit peak. \\
$^e$ Offset indicates the offset of the supernova in arcseconds from the coordinates of the host nucleus, taken from NED. \\
$^f$ Discovery ages are given in days relative to peak. All ages are approximate and are only listed if a clear age was given in the classification telegram. \\
$^g$ Several host names have been abbreviated due to space constraints. \\
\vspace{-0.5cm}
\end{minipage}
\end{table}


\begin{table}
\begin{minipage}{\textwidth}
\bigskip\bigskip
\centering
\fontsize{7}{8}\selectfont
\caption{Non-ASAS-SN Supernovae}
\label{table:other_sne}
\begin{tabular}{@{}l@{\hspace{0.15cm}}l@{\hspace{0.15cm}}c@{\hspace{0.15cm}}c@{\hspace{0.15cm}}c@{\hspace{0.15cm}}l@{\hspace{0.15cm}}c@{\hspace{0.15cm}}c@{\hspace{0.15cm}}c@{\hspace{0.15cm}}c@{\hspace{0.15cm}}c@{\hspace{0.15cm}}l@{\hspace{0.15cm}}c@{\hspace{0.15cm}}c} 
\hline
\vspace{-0.14cm}
 & & & & & & & & & & & & & \\
 & IAU & Discovery &  & & & & & & Offset & & & & \\
 SN Name & Name & Date & RA$^a$ & Dec.$^a$ & Redshift & $m_{peak}^b$ & $V_{peak}^{c,d}$ & $g_{peak}^{c,d}$ & (arcsec)$^e$ & Type & Host Name & Discovered By$^f$ & Recovered?$^g$ \\
\vspace{-0.23cm} \\
\hline
\vspace{-0.17cm}
 & & & & & & & & & & & \\
2018K & SN 2018K & 2018-01-03.13 & 01:50:50.48 & +48:21:13.70 & 0.02350 & 17.2 & --- & --- & 22.14 & Ia-91T & UGC 01303 & Amateurs & No \\ 
2018ec & SN 2018ec & 2018-01-03.40 & 10:27:50.77 & -43:54:06.30 & 0.00935 & 15.1 & 16.0* & --- & 9.01 & Ic & NGC 3256 & Amateurs & No \\ 
2018bi & SN 2018bi & 2018-01-05.29 & 02:19:53.28 & +29:02:02.70 & 0.01663 & 17.1 & --- & --- & 10.71 & Ia & UGC 01792 & Amateurs & No \\ 
ATLAS18eaa & SN 2018dv & 2018-01-05.35 & 02:32:01.03 & +08:35:16.15 & 0.03069 & 16.9 & --- & 16.8 & 21.20 & Ia & WISEA J023202.09+083530.7 & ATLAS & No \\ 
Gaia18abp & SN 2018bl & 2018-01-06.44 & 08:24:11.59 & -77:47:16.55 & 0.01782 & 16.8 & --- & 16.9* & 22.56 & II & ESO 018- G 009 & Gaia & Yes \\ 
Gaia18acg & SN 2018dz & 2018-01-08.70 & 01:12:05.06 & -69:04:59.93 & 0.02000 & 17.2 & --- & 17.2 & 0.92 & II & WISEA J011205.13-690459.7 & Gaia & No \\ 
PS18ej & SN 2018fy & 2018-01-10.51 & 09:53:20.04 & -18:25:09.30 & 0.01200 & 17.3 & 16.8* & --- & 11.70 & II & ESO 566- G 023 & Pan-STARRS & No \\ 
ATLAS20jns & SN 2018lrd & 2018-01-10.61 & 11:28:30.41 & +58:33:44.04 & 0.01041 & 17.1 & --- & --- & 14.80 & Ib & NGC 3690 & ATLAS & No \\ 
2018gj & SN 2018gj & 2018-01-12.24 & 16:32:02.30 & +78:12:40.94 & 0.00454 & 14.4 & 14.5* & 14.8 & 122.76 & II & NGC 6217 & Amateurs & Yes \\ 
ATLAS18ebo & SN 2018jj & 2018-01-14.27 & 00:58:28.10 & -05:52:32.97 & 0.03800 & 17.0 & --- & 16.8 & 0.37 & Ia & PSO J014.6171-05.8759 & ATLAS & No \\ 
Gaia18adx & SN 2018hi & 2018-01-15.16 & 08:53:40.57 & -25:03:32.04 & 0.02539 & 17.9 & 17.1* & --- & 13.80 & II & WISEA J085339.57-250330.0 & Gaia & No \\ 
2018gv & SN 2018gv & 2018-01-15.68 & 08:05:34.61 & -11:26:16.30 & 0.00527 & 12.8 & 13.0* & --- & 63.74 & Ia & NGC 2525 & Amateurs & Yes \\ 
ATLAS18ecc & SN 2018kc & 2018-01-17.59 & 10:30:58.44 & +23:47:18.25 & 0.06369 & 17.6 & --- & 17.5* & 11.40 & Ia & WISEA J103059.27+234718.8 & ATLAS & No \\ 
ATLAS18ebx & SN 2018ke & 2018-01-17.63 & 13:08:39.53 & -41:27:16.04 & 0.01043 & 17.7 & --- & 17.7* & 28.10 & II & ESO 323- G 085 & ATLAS & No \\ 
kait-18A & SN 2018ie & 2018-01-18.50 & 10:54:01.06 & -16:01:21.40 & 0.01423 & 16.6 & --- & 16.9* & 37.28 & Ic & NGC 3456 & LOSS & Yes \\ 
\vspace{-0.22cm}
 & & & & & & & & & & & \\
\hline
\end{tabular}
\smallskip
\\
\raggedright
\noindent This table is available in its entirety in a machine-readable form in the online journal. A portion is shown here for guidance regarding its form and content.\\
$^a$ Right ascension and declination are given in the J2000 epoch. \\
$^b$ Magnitudes are taken from D.~W.~Bishop's Latest Supernova website, as described in the text, and may be from different filters. \\
$^c$ All $V-$ and $g-$band peak magnitudes are measured from ASAS-SN data for cases where the supernova was detected. \\
$^d$ Magnitudes marked with a ``*'' are derived from maximum detected magnitude rather than a fit peak. \\
$^e$ Offset indicates the offset of the supernovae in arcseconds from the coordinates of the host nucleus, taken from NED. \\
$^f$ ``Amateurs'' indicates discovery by any number of non-professional astronomers, as described in the text. \\
$^g$ Indicates whether the supernova was independently recovered in ASAS-SN data or not.
\end{minipage}
\vspace{-0.5cm}
\end{table}

\end{landscape}
\pagebreak
\begin{landscape}


\begin{table}
\begin{minipage}{\textwidth}
\centering
\fontsize{7}{8}\selectfont
\caption{ASAS-SN Supernova Host Galaxies}
\label{table:asassn_hosts}
\begin{tabular}{@{}l@{\hspace{0.15cm}}l@{\hspace{0.15cm}}c@{\hspace{0.15cm}}c@{\hspace{0.15cm}}c@{\hspace{0.15cm}}c@{\hspace{0.15cm}}c@{\hspace{0.15cm}}c@{\hspace{0.15cm}}c@{\hspace{0.15cm}}c@{\hspace{0.15cm}}c@{\hspace{0.15cm}}c@{\hspace{0.15cm}}c@{\hspace{0.15cm}}c@{\hspace{0.15cm}}c@{\hspace{0.15cm}}c@{\hspace{0.15cm}}c@{\hspace{0.15cm}}c} 
\hline
\vspace{-0.14cm}
 & & & & & & & & & & & & & & \\
 & & SN & SN & SN Offset & & & & & & & & & & \\
Galaxy Name & Redshift & Name & Type & (arcsec) & $A_V^a$ & $NUV^b$ & $u^c$ & $g^d$ & $r^d$ & $i^d$ & $z^d$ & $y^d$ & $J^e$ & $H^e$ & $K_S^{e,f}$ & $W1^g$ & $W2^g$ \\ 
\vspace{-0.23cm} \\
\hline
\vspace{-0.17cm}
 & & & & & & & & & & & & & & \\
2MASX J03075327-4544235 & 0.06281 & ASASSN-18ae & Ia & 7.52 & 0.000 & 20.90 0.26 & --- & --- & --- & --- & --- & --- & 13.94 0.05 & 13.19 0.05 & 12.89 0.07 & 12.90 0.03 & 12.92 0.03 \\ 
GALEXMSC J051147.86-401142.5 & --- & ASASSN-18aa & Ia & 0.84 & 0.101 & --- & --- & --- & --- & --- & --- & --- & $>$17.0 & $>$16.4 & 16.16 0.06* & 16.59 0.06 & 16.58 0.16 \\ 
LCRS B110329.3-121524 & 0.02563 & ASASSN-18ac & Ia & 0.57 & 0.105 & 21.95 0.40 & --- & 16.24 0.00 & 15.56 0.00 & 15.21 0.00 & 15.02 0.00 & 14.86 0.00 & 13.95 0.06 & 13.28 0.07 & 12.97 0.10 & 13.52 0.03 & 13.51 0.04 \\ 
2MASX J03232113-2207024 & --- & ASASSN-18af & Ia & 6.23 & 0.033 & 19.01 0.06 & --- & 16.64 0.00 & 16.02 0.00 & 15.67 0.00 & 15.46 0.00 & 15.20 0.00 & 14.27 0.07 & 13.55 0.09 & 13.10 0.11 & 13.11 0.02 & 13.06 0.03 \\ 
MCG +07-30-065 & 0.03166 & ASASSN-18ag & Ia & 4.17 & 0.000 & 18.41 0.05 & 17.26 0.02 & 16.49 0.00 & 14.79 0.00 & 14.45 0.00 & 14.19 0.00 & 14.02 0.00 & 12.80 0.04 & 12.06 0.05 & 11.68 0.07 & 11.79 0.02 & 11.56 0.02 \\ 
SDSS J091612.25+390341.8 & --- & ASASSN-18aj & Ia & 1.06 & 0.000 & 22.20 0.39 & 21.81 0.21 & 20.98 0.04 & 20.54 0.03 & 20.30 0.03 & 20.35 0.05 & 21.41 0.22 & $>$17.0 & $>$16.4 & $>$15.6 & --- & --- \\ 
IC 2777 & 0.03993 & ASASSN-18al & Ia & 3.16 & 0.070 & 17.84 0.03 & 16.39 0.01 & 15.24 0.00 & 14.69 0.00 & 14.46 0.00 & 14.28 0.00 & 14.03 0.00 & 13.11 0.04 & 12.40 0.05 & 12.12 0.08 & 11.94 0.02 & 11.80 0.02 \\ 
NGC 3070 & 0.01778 & ASASSN-18an & Ia & 12.50 & 0.053 & 18.26 0.07 & 14.70 0.00 & 13.10 0.00 & 12.45 0.00 & 12.05 0.00 & 11.82 0.00 & 11.56 0.00 & 10.34 0.02 & 9.64 0.02 & 9.39 0.03 & 10.03 0.02 & 10.04 0.02 \\ 
WISE J163554.27+400151.8 & 0.03101 & ASASSN-18am & II & 8.55 & 0.000 & 19.48 0.07 & 18.76 0.03 & 17.78 0.00 & 17.49 0.00 & 17.29 0.00 & 17.26 0.01 & 17.14 0.01 & $>$17.0 & $>$16.4 & 14.73 0.03* & 15.16 0.03 & 15.05 0.06 \\ 
KUG 0143+322 & 0.03750 & ASASSN-18ap & II & 1.79 & 0.085 & 17.91 0.04 & 16.79 0.01 & 15.51 0.00 & 14.95 0.00 & 14.64 0.00 & 14.48 0.00 & 14.25 0.00 & 13.24 0.04 & 12.53 0.05 & 12.27 0.07 & 12.34 0.02 & 12.23 0.02 \\ 
SDSS J102119.17+205436.5 & --- & ASASSN-18ao & Ia & 0.77 & 0.077 & --- & 20.29 0.11 & 19.66 0.02 & 19.35 0.02 & 19.16 0.02 & 19.25 0.03 & 19.04 0.05 & $>$17.0 & $>$16.4 & 16.65 0.13* & 17.08 0.13 & 17.27 --- \\ 
2MASX J03301134-1331226 & 0.04098 & ASASSN-18ar & Ia & 3.00 & 0.149 & 21.22 0.29 & --- & 15.38 0.00 & 14.56 0.00 & 14.11 0.00 & 13.91 0.00 & 13.70 0.00 & 12.57 0.03 & 11.84 0.04 & 11.51 0.06 & 11.82 0.02 & 11.83 0.02 \\ 
GALEXASC J050811.97-543842.4 & --- & ASASSN-18ba & Ia & 1.50 & 0.325 & 20.04 0.18 & --- & --- & --- & --- & --- & --- & $>$17.0 & $>$16.4 & 14.98 0.03* & 15.41 0.03 & 15.26 0.05 \\ 
2MASX J14211713-0637416 & --- & ASASSN-18az & Ia & 6.05 & 0.109 & 17.69 0.05 & --- & 15.82 0.00 & 15.32 0.00 & 15.04 0.00 & 14.87 0.00 & 14.65 0.00 & 13.84 0.06 & 13.16 0.08 & 12.68 0.09 & 13.27 0.03 & 13.27 0.04 \\ 
KUG 0921+406 & 0.02779 & ASASSN-18bj & Ic-BL & 7.45 & 0.009 & 16.43 0.02 & 15.95 0.01 & 17.32 0.00 & 16.51 0.00 & 14.98 0.00 & 14.86 0.00 & 14.64 0.00 & 13.72 0.05 & 12.99 0.07 & 12.70 0.08 & 12.66 0.02 & 12.50 0.03 \\ 

\vspace{-0.22cm}
 & & & & & & & & & & & & & & \\
\hline
\end{tabular}
\smallskip
\\
\raggedright
\noindent This table is available in its entirety in a machine-readable form in the online journal. A portion is shown here for guidance regarding its form and content. Uncertainty is given for all magnitudes, and in some cases is equal to zero.\\
$^a$ Galactic extinction taken from \citet{schlafly11}. \\
$^b$ No magnitude is listed for those galaxies not detected in GALEX survey data. \\
$^c$ No magnitude is listed for those galaxies not detected in SDSS survey data. \\
$^d$ No magnitude is listed for those galaxies not detected in Pan-STARRS survey data. \\
$^e$ For those galaxies not detected in 2MASS data, we assume an upper limit of the faintest galaxy detected in each band from our sample. \\
$^f$ $K_S$-band magnitudes marked with a ``*'' indicate those estimated from the WISE $W1$-band data, as described in the text. \\
$^g$ No magnitude is listed for those galaxies not detected in AllWISE survey data. \\
\end{minipage}
\vspace{-0.5cm}
\end{table}


\begin{table}
\begin{minipage}{\textwidth}
\bigskip\bigskip
\centering
\fontsize{7}{8}\selectfont
\caption{Non-ASAS-SN Supernova Host Galaxies}
\label{table:other_hosts}
\begin{tabular}{@{}l@{\hspace{0.15cm}}l@{\hspace{0.15cm}}c@{\hspace{0.15cm}}c@{\hspace{0.15cm}}c@{\hspace{0.15cm}}c@{\hspace{0.15cm}}c@{\hspace{0.15cm}}c@{\hspace{0.15cm}}c@{\hspace{0.15cm}}c@{\hspace{0.15cm}}c@{\hspace{0.15cm}}c@{\hspace{0.15cm}}c@{\hspace{0.15cm}}c@{\hspace{0.15cm}}c@{\hspace{0.15cm}}c@{\hspace{0.15cm}}c@{\hspace{0.15cm}}c} 
\hline
\vspace{-0.14cm}
 & & & & & & & & & & & & & & \\
 & & SN & SN & SN Offset & & & & & & & & & & \\
Galaxy Name & Redshift & Name & Type & (arcsec) & $A_V^a$ & $NUV^b$ & $u^c$ & $g^d$ & $r^d$ & $i^d$ & $z^d$ & $y^d$ & $J^e$ & $H^e$ & $K_S^{e,f}$ & $W1^g$ & $W2^g$ \\ 
\vspace{-0.23cm} \\
\hline
\vspace{-0.17cm}
 & & & & & & & & & & & & & & \\
UGC 01303 & 0.02350 & 2018K & Ia-91T & 22.14 & 0.897 & --- & --- & 14.95 0.00 & 13.66 0.00 & 13.08 0.00 & 12.92 0.00 & 12.81 0.00 & 11.42 0.03 & 10.72 0.05 & 10.39 0.04 & 11.30 0.02 & 11.34 0.02 \\ 
NGC 3256 & 0.00935 & 2018ec & Ic & 9.01 & 0.545 & --- & --- & --- & --- & --- & --- & --- & 9.29 0.01 & 8.55 0.01 & 8.17 0.02 & 8.33 0.02 & 7.54 0.02 \\ 
UGC 01792 & 0.01663 & 2018bi & Ia & 10.71 & 0.309 & --- & 15.84 0.02 & 14.30 0.00 & 13.48 0.00 & 13.03 0.00 & 12.91 0.00 & 12.83 0.00 & 11.70 0.03 & 11.00 0.04 & 10.71 0.06 & 12.01 0.03 & 11.96 0.03 \\ 
WISEA J023202.09+083530.7 & 0.03069 & ATLAS18eaa & Ia & 21.20 & 0.321 & --- & --- & 15.48 0.00 & 14.59 0.00 & 14.11 0.00 & 13.99 0.00 & 13.58 0.00 & 12.47 0.03 & 11.69 0.04 & 11.41 0.06 & 11.73 0.02 & 11.81 0.02 \\ 
ESO 018- G 009 & 0.01782 & Gaia18abp & II & 22.56 & 0.385 & --- & --- & --- & --- & --- & --- & --- & 11.97 0.04 & 11.28 0.06 & 10.98 0.09 & 11.00 0.02 & 9.76 0.02 \\ 
WISEA J011205.13-690459.7 & --- & Gaia18acg & II & 0.92 & 0.009 & --- & --- & --- & --- & --- & --- & --- & $>$17.0 & $>$16.4 & 16.22 0.06* & 16.65 0.06 & 16.70 0.20 \\ 
ESO 566- G 023 & --- & PS18ej & II & 11.70 & 0.117 & 18.10 0.04 & --- & 16.03 0.00 & 15.42 0.00 & 15.26 0.00 & 15.02 0.00 & 15.09 0.01 & 13.97 0.07 & 13.33 0.07 & 13.06 0.13 & 13.58 0.03 & 13.45 0.03 \\ 
NGC 3690 & 0.01041 & ATLAS20jns & Ib & 14.80 & 0.000 & --- & --- & 13.76 0.00 & 12.91 0.00 & 12.76 0.00 & 17.63 0.00 & 17.46 0.01 & $>$17.0 & $>$16.4 & $>$15.6 & --- & --- \\ 
NGC 6217 & 0.00454 & 2018gj & II & 122.76 & 0.137 & --- & 13.91 0.00 & 14.12 0.00 & 13.39 0.00 & 11.97 0.00 & 12.16 0.00 & 12.62 0.00 & 9.76 0.02 & 9.08 0.02 & 8.81 0.02 & 9.85 0.02 & 9.53 0.02 \\ 
PSO J014.6171-05.8759 & --- & ATLAS18ebo & Ia & 0.37 & 0.256 & --- & 25.05 1.64 & 22.43 0.12 & 22.61 0.12 & 22.30 0.11 & 22.08 0.20 & --- & $>$17.0 & $>$16.4 & $>$15.6 & --- & --- \\ 
WISEA J085339.57-250330.0 & 0.02539 & Gaia18adx & II & 13.80 & 0.573 & 18.87 0.12 & --- & 15.21 0.00 & 14.50 0.00 & 14.00 0.00 & 13.86 0.00 & 13.75 0.00 & 12.32 0.04 & 11.60 0.04 & 11.29 0.07 & 11.49 0.02 & 11.38 0.02 \\ 
NGC 2525 & 0.00527 & 2018gv & Ia & 63.74 & 0.373 & 14.14 0.01 & --- & 12.68 0.00 & 11.81 0.00 & 13.13 0.00 & 14.48 0.00 & 12.90 0.00 & 9.75 0.02 & 9.13 0.03 & 8.83 0.05 & 10.94 0.02 & 10.77 0.02 \\ 
WISEA J103059.27+234718.8 & 0.06369 & ATLAS18ecc & Ia & 11.40 & 0.000 & 21.00 0.26 & 18.20 0.03 & 16.23 0.00 & 15.35 0.00 & 14.87 0.00 & 14.77 0.00 & 14.61 0.00 & 13.59 0.04 & 13.02 0.07 & 12.58 0.07 & 12.73 0.02 & 12.68 0.03 \\ 
ESO 323- G 085 & 0.01043 & ATLAS18ebx & II & 28.10 & 0.369 & 16.57 0.02 & --- & --- & --- & --- & --- & --- & 12.73 0.05 & 12.12 0.07 & 11.88 0.10 & 12.59 0.02 & 12.50 0.02 \\ 
NGC 3456 & 0.01423 & kait-18A & Ic & 37.28 & 0.161 & 15.45 0.01 & --- & 15.13 0.00 & 18.02 0.01 & 12.59 0.00 & 12.86 0.00 & 17.41 0.01 & 11.11 0.03 & 10.41 0.03 & 10.12 0.05 & 11.29 0.02 & 11.18 0.02 \\ 

\vspace{-0.22cm}
 & & & & & & & & & & & & & & \\
\hline
\end{tabular}
\smallskip
\\
\raggedright
\noindent This table is available in its entirety in a machine-readable form in the online journal. A portion is shown here for guidance regarding its form and content. Uncertainty is given for all magnitudes, and in some cases is zero. \\
$^a$ Galactic extinction taken from \citet{schlafly11}. \\
$^b$ No magnitude is listed for those galaxies not detected in GALEX survey data. \\
$^c$ No magnitude is listed for those galaxies not detected in SDSS survey data. \\
$^d$ No magnitude is listed for those galaxies not detected in Pan-STARRS survey data. \\
$^e$ For those galaxies not detected in 2MASS data, we assume an upper limit of the faintest galaxy detected in each band from our sample. \\
$^f$ $K_S$-band magnitudes marked with a ``*'' indicate those estimated from the WISE $W1$-band data, as described in the text. \\
$^g$ No magnitude is listed for those galaxies not detected in AllWISE survey data. \\
\end{minipage}
\vspace{-0.5cm}
\end{table}

\end{landscape}


\end{document}